\documentclass[UTF8,a4paper]{article}
\usepackage{amsmath}
\usepackage{graphicx}
\usepackage{subfigure}
\usepackage{epstopdf}
\usepackage{geometry}
\usepackage{ulem}
\usepackage{indentfirst}
\usepackage{amsfonts,amssymb,dsfont}
\usepackage{setspace}
\usepackage{threeparttable}
\usepackage{amsmath,mathtools,amsthm}
\usepackage{array}
\usepackage{extarrows}
\usepackage{upgreek}

\makeatletter
\renewcommand*\env@matrix[1][\arraystretch]{%
  \edef\arraystretch{#1}%
  \hskip -\arraycolsep
  \let\@ifnextchar\new@ifnextchar
  \array{*\c@MaxMatrixCols c}}
\makeatother
\begin{document}


\title{\bf Attractive polaron formed in doped nonchiral/chiral parabolic system within ladder approximation}
\author{Chen-Huan Wu
\thanks{chenhuanwu1@gmail.com}
\\College of Physics and Electronic Engineering, Northwest Normal University, Lanzhou 730070, China}

\maketitle
\vspace{-30pt}
\begin{abstract}
\begin{large}

We investigate the properties of attractive polaron formed by a single impurity dressed with the particle-hole excitations
in a three-dimensional (3D) doped (extrinsic) parabolic system.
Base on the single particle-hole variational ansatz,
we study the pair propagator, self-energy, and the non-self-consistent medium $T$-matrix.
The non-self-consistent $T$-matrix discussed in this paper 
contains only the open channel 
since we do not consider the shift of center-of-mass due to the resonance (e.g., induced by the magnetic field).
The scattering form factor is discussed in detail for the chiral case and compared to the non-chiral one.
The effects of the bare coupling strength, which is momentum-cutoff-dependent, are also discussed.
We found that the pair propagator and the related quantities, like the self-energy, spectral function, induced effective mass,
and residue (spectral weight), all exhibit some novel features in the low-momentum regime 
which are different from the high one.
That also related to the polaronic instabilities as well as the many-body fluctuation and nonadiabatic/adiabatic dynamics.
The effects of chirality and the finite-temperature (much lower than the critical one)
are studied in detail, and our results can also be used in exploring other weak-coupling systems
with mobile impurity.\\
\\
$ Keywords$: Attractive polaron;
non-self-consistent medium $T$-matrix;
Retarded self-energy;
Pair propagator;
Chiral system;
Ladder approximation\\

\end{large}

\end{abstract}
\begin{large}

\section{Introduction}

The polarons formed in atomic lattice system\cite{Grusdt F,amacho-Guardian A} or 
crystal lattice system have been intensely studied,
and both the theoretical and experimental efforts on the polaronic dynamics
have been made to explore the pairing instability\cite{Yi W},
phase transition\cite{Onofrio R,Sowinski T,Kagan M Y}, 
and the thermal excited collective modes\cite{Guenther N E}.
Besides, a wide range of experimental tools, like the Feshbach resonance and radio-frequency spectroscopy
\cite{Hu M G,Massignan P3,Veillette M},
have been applied in detecting the formation of polarons,
which reveal great potential of polaron in superconductivity and quantum simulation.
The theoretical tools, like the variational approach\cite{Sidler M,scripta}, 
perturbation theory\cite{Christensen R S},
and diagrammatic Monte Carlo\cite{Field B,Prokof'ev N,
Marchand D J J,Sous},
are also widely applied in studying the polaronic dynamics in both the strong-coupling and 
weak coupling regimes\cite{Marchand D J J,Guenther N E,biexcitons}.
However, the properties of polaron formed in a topological or chiral systems,
especially at finite temperature and with the thermal-excitations, have rarely been studied.
In this paper, we focus on the properties of Fermion polaron formed in
a three-dimensional (3D) doped parabolic system in a crystal lattice model,
where effects of chirality and the finite-temperature are studied in detail.

Consider the effect of impurities (in coherent background),
the polaron as an excited quasiparticle in the population/spin-imbalanced Fermi gases, BEC, or the topological insulator
\cite{amacho-Guardian A,Shvonski A,Qin F},
are important when the many-body effect is taken into account.
Unlike the gapless Dirac system where the electron-electron interaction usually gives rise to the renormalization of quasiparticle
velocity\cite{Elias D C,para},
for systems with quadratic dispersion, the electron-electron interaction usually gives rise to the effective mass renormalization\cite{Sarma S D},
which will be discussed in this paper,
and we note that the effective mass approximation\cite{Devreese J T,Parish M M,Christensen R S}
 is applicable for the polaron in small momentum regime.
Besides, since the spin rotation is missing in our system 
in the presence of $\delta$-type impurity field,
the spin structure is fixed,
and the interacting spins between impurity and majority particle are usually opposite in direction (due to the Pauli principle)
which provides the opportunity to form the Cooper pair and the strongly bound dimer in superfluid.
That also implies the low-temperature environment with weaker spin relaxation is beneficial 
to the formation of polarons,
that can also be seen from the finite temperature pair propagator as presented in this paper.
We present in Appendix.A a brief discussion of the variational approach in mean-field approximation
which is valid in the weak interacting regime with non-too-low density\cite{?wik J A}
for the polaron dressed by the partially polarized excitations-cloud,
however, the mean-field approximation sometimes overestimates the interaction effect\cite{Li W}
in strongly interacting regime (with tightly bound dimers) where the self-trapping, soliton, and breather are harder
to formed than that in the weakly bound pairs (e.g., the BCS superfluid state in the non-Fermi-liquid picture).

For ultracold atomic systems,
the stable repulsive polarons are most likely to be found in the side away from the Feshbach resonance (for gases) 
where polaron energy is large and positive (with $1/k_{F}a \gg 1$;
$a_{\psi\phi}$ is the scattering length related to the impurity-majority interaction,
which is produced by the attractive potential and $k_{F}$ is the Fermi wave vector),
while
in strong scattering region ($1/k_{F}a\lesssim 1$),
the tightly-bound molecule (within Fermi-liquid picture) could also be found with binding energy 
$E_{b}=\hbar^{2}/2a^{2}\widetilde{m}>0$\cite{Parish M M,Koschorreck M},
as experimentally realized in Ref.\cite{Scazza F,J?rgensen N B}.
However, in most cases the repulsive polaron is thermodynamically unstable in the 
strong interacting regiem
even at low-temperature.
  While for attractive polaron,
  whose eigenenergy is negative,
it can be observed in the 
solid state systems\cite{Koschorreck M,Mogulkoc A,Modarresi M,Ding Z H,Mogulkoc A2,Falck J P} 
including the topological materials,
through the, e.g., substrate-related polaronic effect.
Besides, the effects of both the intrinsic and extrinsic (like the electric field-induced Rashba) spin-orbit coupling 
on the polaron
have also been explored\cite{Mogulkoc A,Modarresi M,Ding Z H,Mogulkoc A2},
and it is found that the spin-orbit coupling is also related to the polaron-molecule transition\cite{Yi W}.
The quasiparticle residue $Z$ (spectral weight), as an important property of the polarons,
which can be measured by the Rabi oscillations\cite{Kohstall C},
is finite even at zero-energy state for the parabolic system or normal Dirac semimetals,
but it vanishes at zero-energy state for the Weyl semimetals which has higher-order dispersion\cite{Wang J R,highero}
(multi-Weyl semimetal).

We focus on the weak-coupling (and short-range) regime in nonadiabatic configuration throughout, 
where the quasiparticle spectral weight 
is away from zero and the polaron is well-defined (and thus long-lived),
since both the strong interaction\cite{Werman Y} and high temperature\cite{Guenther N E} will lead to ill-defined polaron (with large imaginary part of self-energy).
While the long-range Coulomb interaction, which will gives rise to adiabatic effect
and thus accelerates the electrons above the Fermi surface
(by logarithmically lowers the density-of-states at zero energy), is ignored here due to the screening effect.
For example, by including a Thomas-Fermi screening wave vector ($k_{TF}=me^{2}/\epsilon$)
 into the Coulomb interaction (i.e., the static part of the particle-hole bubble),
the interaction range will become shorter,
and satisfies $a_{B}k_{F}\ll 1$\cite{Miserev D,Sidler M} ($a_{B}=k_{TF}^{-1}$) at small carrier concentration,
i.e., the diluteness condition\cite{Randeria M,Hwang E H3} which implies that the attractive range and the Fermi wave vector as well
as the Fermi energy are small.
In this limit,
the two-body pairing dynamics is completely characterized by the low-energy $T$ matrix\cite{Randeria M}
since the short-range repulsive interaction is very weak.
In two-body problem ($k_{F}\rightarrow 0$),
 the feature of screened polaron will
become more obvious in the non-relativistic limit,
which has a slower motion compared to the intrinsic Dirac electron (relativistic),
and
that also gives rise to the nonadiabatic dynamics.
Due to the nonadiabatic character, the Matsubara Green's function
is fully momentum- and frequency-dependent in renormalization group flow,
which is different from the 
bare Coulomb interaction in the clean Dirac system with the instantaneous nature.
The vertex function as well as the pair propagator is also fully momentum- and frequency-dependent.

For the attractive Fermionic polaron formed through the attractive interaction 
between the impurity with the electron-hole pairs excited in a doped parabolic system in weak coupling regime,
its polaronic effects are investigated by using the non self-consistent $T$-matrix approach 
(consist of the undressed propagators, and thus the particle/energy conserving is broken)
within the ladder approximation.
The accuracy of non-self-consistent method used here has been verified\cite{Rath S P,Holm B}.
Note that, however, the energy of polaron obtained here will show deviation from the DFT result (for unit cell)\cite{Sio W H}
since the Coulomb energy are ignored, also, the non-self-consistent treatment will results in deviations from the real
total energy\cite{Holm B}.
We consider the low-doping case of the parabolic system
so that the interband electron-hole excitation can be created by the impurity,
and not be suppressed by the phase space restriction.
Thus the system discussed in this paper 
will exhibit some Fermi-liquid charateristics,
which can be seen from the power-law behavior in frequency of the self-energy imaginary part.
Our theory can also be applied to the massive (doped) Dirac or Weyl systems
where the band structure in low-energy regime is gapped.

In Sec.2, we introduce the model, where 
the Hamiltonian of the nonchiral or chiral impurity are presented.
In Sec.3, we present the expressions of the non-self-consistent $T$-matrix, part-propagator, and self-energy.
The interacting Hamiltonian are presented, where the polaron problem can be related to the
Schrodinger-type eigenvalue problem subjected to a wave functional constraint,
similar to Ref.\cite{Sio W H,Yu Z Q}.
Besides, the validity for non-self-consistent $T$-matrix in studying the many-body effect is also discussed in this section.
In Sec.4, we present the main results of this paper,
where the single-channel BCS model as well as the variational wave function are introduced,
and the pair propagator-related quantities (including the self-energy and spectral function)
 are derived for the chiral and non-chiral cases
for zero-temperature case.
The induced
 effective masses and the residue are also calculated for different coupling strength.
The scattering form factor here is different from the one with single-particle propagation.
The polaronic stability and the many-body fluctuation are also discussed.
In Sec.5, we discuss the pair propagator
and the relaxation time at finite temperature (but much lower than the critical one)  
 which exhibit some differences compared to
the zero-temperature case.
In Appendix.A, we study the variational approach in mean-field approximation.
In Appendix.B, we further discuss the diagrammatic approach of the polaron self-energy
within extended ladder approximation (to arbitrary order of bare coupling $g_{b}$, but it is different from
the perturbative expression as shown in Ref.\cite{Christensen R S}).
In Appendix.C, we take the electron-phonon coupling (due to the thermal excitations) into account,
and further discuss the Chevy Ansatz of the three-body polaron (including a finite number of phonons).

\section{Theories}

The properties of the mobile impurities are widely studied\cite{Koschorreck M,Scazza F,Fratini E}, 
which unlike the 
immobile one: without the Kondo effect and indirect exchange interactions.
For such mobile impurity,
 the $\delta$-type impurity field 
(within Born approximation) is valid in measuring the effect of impurity-interaction.
However, that also cause the vanishing of the spin rotation as well as the intrinsic spin current during the impurity-majority 
scattering (collision) process.
Then, the fixed spin structure guarantees that there exists only the singlet pairing between the impurity and majority particles,
otherwise the triplet pairing exists as discussed in Ref.\cite{Yi W}.
In Fermi gases or the dilute BEC, the interactions between impurity and majority component, and that
between the majority particles needed to taken into account,
such problems are usually treated by the non-self-consistent many-body $T$-matrix (i.e., the ladder approximation).
Here we consider the interactions between the Fermions (bath) and the single Fermionic impurity.
The difference compared to the normal solid state system\cite{Jian S K,Yang B J,Han S E} is that we are
taking the mobile impurity into account and 
note that we still focus on the single impurity problem since the
Fermi polaron is well defined (with symmetry and easy-to-identify spectral function) in the single-impurity limit\cite{Koschorreck M}.
At first, we write the microscopic Euclidean action as
\begin{equation} 
\begin{aligned}
S=\int d\tau d^{3}r \{\psi^{\dag}[\partial_{\tau}+H_{0}(k)]\psi+\frac{1}{2}\sum_{\alpha=x,y,z}(\partial_{\alpha}\phi)^{2}
+\frac{1}{2}g_{\psi\psi}\psi^{\dag}\psi\psi^{\dag}\psi+g_{\psi\phi}\psi^{\dag}\psi\phi^{\dag}\phi\},
\end{aligned}
\end{equation}
where $\psi$ and $\phi$ denote the majority and impurity field, respectively,
$g_{\psi\psi}$ and $g_{\psi\phi}$ are the intraspecies (only around the impurity) and interspecies coupling, respectively,
for the mobile particles.
For parabolic chiral system (doped),
the noninteracting Hamiltonian of the impurity reads
\begin{equation} 
\begin{aligned}
H_{0}=&\frac{|{\bf p}|^{2}}{2m}\hat{{\bf p}}\cdot {\pmb \sigma}+\hbar v_{z}p_{z}\sigma_{z}-\mu,\\
=&\frac{p_{x}^{2}}{2m}\sigma_{x}+2\frac{p_{x}p_{y}}{2m}\sigma_{y}-\frac{p_{y}^{2}}{2m}\sigma_{x}+\hbar v_{z}p_{z}\sigma_{z}-\mu.
\end{aligned}
\end{equation}
Note that here $p$ can be replaced by other momenta to represent the other particles (like the majority component).
While for the nonchiral parabolicn system, the Hamiltonian can be obtain by removing the term $\hat{{\bf p}}\cdot {\pmb \sigma}$
in above equation or replacing it by a (spin) Pauli matrix $\sigma_{z}$.
We assume that the longitudinal term involving $p_{z}$ is small, and thus the eigenenergy can be approximated as $\frac{p^{2}}{2m}-\mu$.
Due to the chirality (as can be seen, the momentum is locked with the Pauli matrices acting on the band space),
the eigenvectors can be written as\cite{highero} (after ignoring the $p_{z}$-term)
\begin{equation} 
\begin{aligned}
|p\rangle=&\frac{e^{i{\bf p}\cdot {\bf r}}}{\sqrt{S}}\frac{1}{\sqrt{2}}
\begin{pmatrix}
1\\
\lambda e^{i2\phi\lambda}
\end{pmatrix}\\
=&\frac{e^{i{\bf p}\cdot {\bf r}}}{\sqrt{S}}\frac{1}{\sqrt{2}}
\begin{pmatrix}
e^{-i\phi}\\
\lambda e^{i\phi}
\end{pmatrix},\\
\end{aligned}
\end{equation}
where 
$\phi ={\rm arctan}\frac{p_{y}}{p_{x}}$
is the polar angle of ${\bf p}$.
$\lambda=\pm 1$ correspond to the electron and hole states, respectively.

In this paper, we only consider the interspecies interaction,
$g_{\psi\phi}$, which
describes the bare attractive contact interaction strength.
The contact potential (with Gaussian broadening)
is used here since the size of pair is much larger than the range of potential,
thus the long-range Coulomb repulsion is reduced. 
For relativistic particles in, e.g., the 3D Dirac/Weyl system,
the mass term is necessary to form the polaron.
We here use the coupling constants of the impurity-Fermion sea mixture system,
$
g_{\psi\psi}^{-1}=\frac{m_{\psi}}{4\pi \hbar^{2}a_{\psi\psi}},
g_{\psi\phi}^{b}=[\frac{\widetilde{m}}{2\pi \hbar^{2}a_{\psi\phi}}-\int\frac{d^{3}k}{(2\pi)^{3}}\frac{2\widetilde{m}}{k^{2}}]^{-1}$,
where 
$\widetilde{m}=m_{\psi}m_{\phi}/(m_{\psi}+m_{\phi})$ is the renormalized mass, superscript $b$ denotes the bare coupling.
 $\Lambda$ is the momentum cutoff.
$a_{\psi\phi}$ is the impurity-majority scattering length.
The $g_{\psi\psi}$ here follows the general definition of the background coupling constant 
which related to the background scattering length $a_{\psi\psi}$.
Indeed, the interaction effect of the polaron system requires the investigation of the effective mass
in contrast to the semimetal system,
especially in the strong interacting regiem
with the obvious renormalization effect and the Fermi-liquid feature.

\section{$T$-matrix and the self-energy}

The non-perturbative ladder approximation 
is applicable not only for the imbalanced Fermi gases or nuclear physics, but also for the solid state systems with finite effective mass.
Further, for large-species Fermion system,
the ladder approximation is similar to the leading order $1/N$-expansion.
In thermodynamic limit with a infinite system 
(i.e., $N\rightarrow\infty$; note that $N$ here is the number of unit cell or the 
quantization volume but not the flavor number),
the polaron self-energy which describes the pairing fluctuation becomes zero\cite{Enss T,Werman Y} 
which implies that the interaction between impurity and majority particles vanishes 
(i.e., without the polaron).
In the strong interacting case,
the self-energy effect as well as the resummation of ladder diagrams (for the forward scattering) are important.
The non-self consistent $T$-matrix
describes the fluctuations in $s$-wave cooper channel.
Firstly, we can write the $T$-matrix between the single impurity and the majority component as
\begin{equation} 
\begin{aligned}
T(p+q,\omega+\Omega)=[\frac{\widetilde{m}}{2\pi \hbar^{2}a_{\psi\phi}}+\Pi(p+q,\omega+\Omega)]^{-1},
\end{aligned}
\end{equation}
where $q$ is the momentum of the majority particle, $p$ is the momentum of the impurity.
Note that for the $T$-matrix here,
we only consider the closed channel scattering (i.e., the bare case),
and without considering the interchanging as well as the spin/valley degrees of freedom.
The term $(p+q)$ can be treated as the center-of-mass momentum.
$\Omega
$ is the Fermionic frequency
since we assume the zero-temperature limit,
similarly,
$\omega
\approx \varepsilon_{p\downarrow}=\frac{\hbar^{2}(p^{2})}{2m_{\phi}}-\mu_{\downarrow}
=\frac{\hbar^{2}(p^{2})}{2m_{\phi}}-{\rm Re}\ \Sigma(p=0,\omega=0)
$ is the Bosonic frequency
where $\Sigma(p,\omega)$ is the impurity self-energy as stated below.
Here $\mu_{\uparrow}\neq \mu_{\downarrow}$ due to the spin-imbalance.
Note that this $T$-matrix is non-self-consistent,
where the bare impurity propagator and the majority propagator as diagrammatically shown by the Bethe-Salpeter equation
(see, e.g., Ref.\cite{polaron2}).
While the self-consistent $T$-matrix requires the dressed impurity propagator which contains the impurity self-energy effect,
and it is more suitable to apply when we take an infinite number of the majority particles into account ,
in which case its statistical properties emerge
including the imbalance between the two majority species\cite{Pietil? V}.
The Bethe-Salpeter equation about the non-self-consistent many-body $T$-matrix reads
\begin{equation} 
\begin{aligned}
T(p+q,&\square;p+q-k')=V_{0}(p+q,\square;p+q-k')\\           
      &+\sum_{k}V_{0}(p+q,\square;k)G^{\phi}_{0}(p+q-k)G^{\psi}_{0}(\square+k)T(p+q-k,\square+k;p+q-k-k')
\end{aligned}
\end{equation}
where $V_{0}$ are the bare impurity-majority interactions,
specially, $V_{0}(p+q,\square;k)$ is the interaction induced by the polarization operator
(consist of the two bare Green's functions; see Appendix. B).
$k,\ k'$ are the relative momentum.
$G^{\psi}_{0}$ and $G^{\phi}_{0}$ are the bare Fermionic and Bosonic Green's function, respectively,
as presented below.
The symbol $\square$ can be omitted, but we retain it here for the integrity of the above equation.
In the absence of the center-of-mass momentum ($p+q=0$),
the Bethe-Salpeter equation is reduced to the Lippmann-Schwinger equation
\begin{equation} 
\begin{aligned}
T(k_{1},k_{2};\omega)=V_{0}(k_{1},k_{2})          
      +\sum_{k_{3}}V_{0}(k_{1},k_{3})\frac{1}{\omega+i0-2\varepsilon_{k_{3}}}T(k_{3},k_{2},\omega).
\end{aligned}
\end{equation}

The impurity-majority pair propagator (unrenormalized) in non-self consistent $T$-matrix approximation reads
\begin{equation} 
\begin{aligned}
\Pi(p+q,\omega+\Omega)
=\int\frac{d^{3}k}{(2\pi)^{3}}\int\frac{d\nu}{2\pi}
                             G_{0}^{\psi}(-\Omega+\nu,k-q)G_{0}^{\phi}(\omega+\Omega-\nu,p+q-k),
\end{aligned}
\end{equation}
where $ G_{0}^{\psi}(-\Omega+\nu,k-q)=[i\nu-i\Omega-\frac{k^{2}}{2m_{\psi}}+\mu_{\uparrow}]^{-1}$ is the noninteracting 
(in the absence of a condensate and the long-range Coulomb interaction) majority particle (Fermion) propagator 
and $G^{\phi}_{0}(\omega+\Omega-\nu,p+q-k)=[i\omega+i\Omega-i\nu-\frac{(p+q-k)^{2}}{2m_{\phi}}+\mu_{\downarrow}]^{-1}$ 
is the bare impurity propagator (not the scalar-field one).
In this majority particle propagator we ignore the perturbation from the single impurity to the Fermi sea.

We at first discuss the self-energy of polaron in the general 
Fermi sea,
which is usually referred to as the polaronic binding energy\cite{Li W} or the molecule binding energy\cite{Yu Z Q,Koschorreck M}.
This self-energy can be obtained by the following impurity-majority interaction Hamiltonian
\begin{equation} 
\begin{aligned}
H_{{\rm int}}=\begin{pmatrix}
g_{\psi\phi}|\Psi_{\psi}|^{2} & 0\\
0& g_{\psi\phi}|\Psi_{\phi}|^{2}
\end{pmatrix},
\end{aligned}
\end{equation}
and the related the interaction energy 
\begin{equation} 
\begin{aligned}
\varepsilon_{\psi\phi}=n_{F}\int d^{3}R \Psi(R)[\frac{-\hbar^{2}\nabla_{R}^{2}}{2\widetilde{m}}+U_{\psi\phi}]\Psi^{\dag}(R),
\end{aligned}
\end{equation}
where $\Psi(R)=(\Psi_{\psi}(R),\Psi_{\phi}(R))$ is the normalized wave function.
$n_{F}=\int\frac{d^{3}q}{(2\pi)^{3}}\int\frac{d\Omega}{2\pi}G_{F}(q,\Omega)e^{i0^{+}\Omega}$ is the numerical density of the Fermions,
where $G_{F}(q,\Omega)$ 
is the dressed (full) Green's function 
and gives the actual Fermion dispersion.
There exist the constrains
\begin{equation} 
\begin{aligned}
4\pi n_{F}\int^{\Lambda}_{0}dR R^{2}\Psi(R)\Psi^{\dag}(R)=&{\bf I},\\
|\Psi(R\ge \Lambda_{r})|=&1,
\end{aligned}
\end{equation}
where ${\bf I}$ is the identity matrix.
Base on the many-body scattering theory at low-temperature,
where we consider only the $s$-wave scattering,
the $T$-matrix in Fermi (or Bose) gases is usually self-consistent,
i.e.,
a two-channel $T$-matrix\cite{Massignan P,Bruun G M}
while in our model,
the $T$-matrix contains only the open channel (i.e., the bare one)
which is non-self-consistent.

To study the many-body effect, the non-self-consistent $T$-matrix is similar but not exactly like the leading order $1/N$ expansion,
since it ignores the dynamical screening effect.
For this reason, the non-self-consistent $T$-matrix is more like the leading-order loop expansion 
within GV approximation rather than the leading order $1/N$ expansion within GW approximation.
It is also found that the static screening (GV) to the Coulomb interaction can be a good approximation for the dynamical screening in the 
low doping regime, especially for carriers frequency of the order of binding energy\cite{Yong C K}.
The related studies are also reported in Refs.\cite{Enss T,Rath S P,Nikoli? P}.
Besides, the validity of the non-self-consistent $T$-matrix in studying the BCS-BEC crossover has also been verified
\cite{Tsuchiya S,Haussmann R}.

In our model,
the self-energy (in second order of interaction) about the interaction between mobile impurity and the bath reads
\begin{equation} 
\begin{aligned}
\Sigma(p,\omega)
&=\sum_{q<k_{F},\Omega}\frac{N_{F}(\Omega)}
{\frac{\widetilde{m}}{2\pi \hbar^{2}a_{\psi\phi}}+\Pi(p+q,\omega+\Omega)},
\end{aligned}
\end{equation}
and the numerator can be replaced by $\theta(k_{F}-q)$ at zero-temperature limit, where $\theta$ stands for the step function.
Note that this self-energy expression describes only the region around the single 
mobile impurity (the attractive polaron).
It is different to the Fermi gases that,
the self-energy of polaron here does not contains the density-dominated term 
as well as the condensate-related spin fluctuation
and the pair propagator contains the chiral factor $F_{\lambda\lambda'}\ (\lambda,\lambda'=\pm 1)$ (the wave function overlap) which 
suppresses the backscattering and is absent in the 2D electron gas.
The chiral factor here is indeed observable in the polarons formed in the Dirac system\cite{Kandemir B S}.
While for 2D electron gas,
$F_{\lambda\lambda'}=1$ and contains only the intraband contribution,
except at a quantum Hall setup with strong magnetic filed as report in Ref.\cite{Bocquillon E}.
We can also see that,
in the narrow gap limit,
the pair propagator reduced to the well known dynamical polarization,
and $T=\Pi^{-1}$.
In the surface of Dirac system,
since away from the condensed phase, the condensate density vanishes but the 
related pairing fluctuations remain as long as 
$g_{\psi\phi}\neq 0$, and the pairing instability exists (especially when the spin-orbit coupling turns on\cite{Yi W})
even in the case of $g_{\psi\psi}=0$.
Further, we note that in 3D the many-body instability (between nonrelativistic particles)
exists as long as the interaction is attractive (since $\int d^{3}r\delta(r)$is converge), 
while the two-body bound state is not required unlike
the 2D case(cf.\cite{Randeria M}).

\section{Pair propagator and related quantities at zero-temperature limit}

To describe the polaronic dynamics, we use the following BCS-type many-body Hamiltonian
\begin{equation} 
\begin{aligned}
H=&\frac{1}{N}\sum_{k}\varepsilon_{k\uparrow}c_{k\uparrow}^{\dag}c_{k\uparrow}+\frac{1}{N}\sum_{p}\varepsilon_{p\downarrow}c_{p\downarrow}^{\dag}c_{p\downarrow}
+\frac{1}{N}\sum_{k,p,q}g_{q}\langle p-q|p\rangle\langle k+q|k\rangle c_{p-q\downarrow}^{\dag}c_{k+q\uparrow}^{\dag}c_{k\uparrow}c_{p\downarrow},
\end{aligned}
\end{equation}
where $N=S/s_{0}$ is the total number of unit cell where $S$ is the total area and $s_{0}$ is the area of unit cell. We omit this factor in the following.

Within one-particle-hole approximation,
which is valid
according to the Monte Carlo calculation and the 
experimental results due to the destructive interference in the presence of more than one particle-hole part,
the variational wave function reads\cite{Combescot R2,Chevy F}
\begin{equation} 
\begin{aligned}
|\psi\rangle_{p}=
\psi_{0}c_{p\downarrow}^{\dag}|0\rangle_{\uparrow}+\sum_{k>k_{F},q<k_{F}}\psi_{kq}c_{p+q-k,\downarrow}^{\dag}
c_{k,\uparrow}^{\dag}c_{q,\uparrow}|0\rangle_{\uparrow},
\end{aligned}
\end{equation}
where $|0\rangle_{\uparrow}=\Pi_{k<k_{F}}c_{k\uparrow}^{\dag}|{\rm vac}\rangle$ is the ground state of the majority particles\cite{Combescot R}
and $|{\rm vac}\rangle$ is the vacuum electron state.
$p$ is the center-of-mass momentum, and $k,q$ are the relative momenta.
We focus on the coherence case, where the masses of impurity and the majority particle are comparable,
and thus decoherence effect\cite{Visuri A M} is weak while the nonadiabatic dynamics is dominating.
The first term in the right-hand-side of above equation describes the free impurity assumed totally delocalized,
$k$ is the momentum of a majority-particle scattered out of Fermi surface, and
$q$ is the momentum of a majority-particle before scattering.
Through the normalization condition $\langle\psi|\psi\rangle=1$, we have,
after minimizing the total energy,
\begin{equation} 
\begin{aligned}
\psi_{kq}=&\psi_{0}\frac{T(p+q,\omega+\Omega)}{\omega-\varepsilon_{p+q-k,\downarrow}-\varepsilon_{k,\uparrow}+\varepsilon_{q,\uparrow}},\\
\psi_{0}=&\frac{1}{\sqrt{1+\sum_{k>k_{F},q<k_{F}}(\frac{\psi_{kq}}{\psi_{0}})^{2}}}.
\end{aligned}
\end{equation}

At zero-temperature limit,
the above renormalized pair propagator can be written as
\begin{equation} 
\begin{aligned}
\Pi(p+q,\omega+\Omega)
=& - \int\frac{d^{3}k}{(2\pi)^{3}}
    \frac{1-N_{F}(\varepsilon_{k\uparrow})}
{-\omega-i0-\Omega+\varepsilon_{k\uparrow}+\varepsilon_{p+q-k\downarrow}}F_{\lambda\lambda'}\\
=&  -\int\frac{d^{3}k}{(2\pi)^{3}}
    \frac{N_{F}(\varepsilon_{k\uparrow})-1}
{\omega+i0+\Omega-\varepsilon_{k\uparrow}-\varepsilon_{p+q-k\downarrow}}F_{\lambda\lambda'}\\
=&-\frac{4\pi}{(2\pi)^{3}}\int^{\Lambda}_{k_{F}}
    \frac{-k^{2}\theta(k-k_{F})}{\omega+i0+\varepsilon_{q\uparrow}-\varepsilon_{k\uparrow}-\varepsilon_{p+q-k\downarrow}}F_{\lambda\lambda'}dk,
\end{aligned}
\end{equation}
where $N_{F}$ is the Fermi-distribution function and it appears only in the presence of nonzero center-of-mass momentum.
Note that instead of using a term $\frac{2\widetilde{m}}{k^{2}}$ to make the pair propagator convergent even in the ultraviolet limit
\cite{Massignan P,Massignan P2}
(renormalized pair propagator),
we here use the momentum cutoff $\Lambda$ similarly to \cite{Parish M M,Sidler M},
which is in the same magnitude of the inversed lattice constant,
and the value of momentum cutoff is setted as 3 eV which is the same as the graphene-like systems\cite{Chae J}.
Interestingly, it is also found that, for open channel $T$-matrix (which is what we focus on throughout this paper),
the open channel shift due to the medium effect (i.e., the integral for open channel propagator over the scattering wave vector
\cite{Bruun G M,Christensen R S}) 
just equals to this term ($\sum_{k}\frac{2\widetilde{m}}{k^{2}}$) in the vacuum limit 
(with zero center-of-mass momentum and zero impurity frequency).
Base on the above expression of the pair propagator, we can obtain that,
the polaron self-energy increases with the increasing mass term or coupling parameter $g_{\psi\phi}$,
however, there is an exception:
when the intrinsic spin-orbit coupling (not the extrinsic one) is presented,
then the increase of $g_{\psi\phi}$ will reduces the self-energy since it will greatly reduces the mass\cite{Mogulkoc A}.
As we can see, although the chiral factor $F_{\lambda\lambda'}$ is contained, 
it has $F_{\lambda\lambda'}=1$ for the non-chiral systems (like the 2D electron gas) or the systems which are
dominated by the backscattering (like the bilayer Dirac system\cite{Adam S}).

Different from the retarded polarization function (density-density correlation) in one-loop approximation
which only describes the scattering of one kind of particle (like the electron) due to the interaction (like the Coulomb interaction),
the pair propagator describes both the scatterings of the impurity and the majority particle (electron-hole pair here).
The second term in above trial wave function corresponds to the excited state where the impurity scattered from state with
intrinsic momentum $p$ to the final state with momentum $p+q-k$, i.e., the scattering wave vector is $q-k$;
while the majority particle (excited electron-hole pairs; whose number
is proportional to the ${\rm Im}\Sigma$)
 scattered from the initial state with momentum $0$ (does not exis yet until
created by the impurity) to the final state with momentum $k-q$, i.e., the scattering wave vector is $k-q$.
Note that here we assume that the scattering wave vectors are small and thus the inter valley scattering can be ignored.
We can see that the scattering wave vectors for impurity and majority particles are opposite in direction.
Thus for such a pair propagator, the spinor wave function overlap (form factor) according to Eq.(3) reads
\begin{equation} 
\begin{aligned}
F_{\lambda\lambda'}
&=\langle (p+q-k)|p\rangle  \langle k-q|0\rangle \\
&=\frac{1}{2}(e^{i\phi_{p+q-k}\lambda'}\ \lambda'e^{-i\phi_{p+q-k}\lambda'})
\begin{pmatrix}
e^{-i\phi_{p}\lambda}\\
\lambda e^{i\phi_{p}\lambda}
\end{pmatrix},
\end{aligned}
\end{equation}
where the $\theta$ is the angle between the initial impurity wave vector $p$ and the scattered wave vector $(q-k)$.
Apparently such form factor is a little different to the one which is widely seen in the chiral solid system\cite{highero,Efimkin D K}.
Through this form factor,
we can clearly see the difference between the pair propagator and the one loop diagrams where the two propagators describe the same particle 
before and after scattering, respectively.
For intraband transition ($\lambda\lambda'=1$; i.e., the initial impurity and the dressed impurity are all located
in the conduction band),
\begin{equation} 
\begin{aligned}
F_{\lambda\lambda'}
&={\rm cos}(\phi_{p+q-k}-\phi_{p})\\
&=\frac{p+(q-k){\rm cos}\theta}{\sqrt{p^{2}+(q-k)^{2}+2p(q-k){\rm cos}\theta}}\\
&\approx 1-\frac{{\rm sin}^{2}\theta}{2p^{2}}(q-k)^{2}.
\end{aligned}
\end{equation}
where $\theta$ is the angle between initial momentum $p$ and the scattering one $q-k$.
For interband transition ($\lambda\lambda'=-1$),
\begin{equation} 
\begin{aligned}
F_{\lambda\lambda'}
&=i{\rm sin}(\phi_{p+q-k}-\phi_{p})\\
&\approx i\sqrt{1-[1-\frac{{\rm sin}^{2}\theta}{2p^{2}}(q-k)^{2}]^{2}}.
\end{aligned}
\end{equation}
Next we only focus on the nonchiral and intraband chiral cases.
For the polarons formed in parabolic systems with nonrelativistic interacting particles,
the renormalized interacting strength can be represented in another representation
\begin{equation} 
\begin{aligned}
g_{\psi\phi}^{-1}=\frac{1}{g_{b}}(1+g_{b}\sum_{k=k_{F}}^{\Lambda}\frac{1}{2\varepsilon_{k}}).
\end{aligned}
\end{equation}
where the bare coupling $g_{b}$ is tunable and it tends to zero
when let $\Lambda\rightarrow\infty$.
Note that for nonrelativistic particles in 3D,
the interaction is independent of the carrier density in ladder diagram\cite{Schick},
and it becomes a constant (momentum independent)
in the limit of zero transferred momentum and zero interaction range,
which corresponds to the first-order approximation in $T$-matrix
in many-body physics: $g_{b}=\langle 0|T|0\rangle$.

Obviously, as showed in Fig.1(a),
the above scattering form factor (Eq.(16)) corresponds to the pair propagator of $\Pi_{11}$,
while for the particle-hole propagator up to second order\cite{Christensen R S} $\Pi_{22}$ as showed in the Fig.1(b), 
we will see that the value of its form factor is the same as the that of $\Pi_{11}$,
i.e.,
\begin{equation} 
\begin{aligned}
F_{\lambda\lambda'}
=\langle (p+q-k)|p\rangle  \langle k-q|0\rangle 
=F'_{\lambda\lambda'}
&=\langle (p+q-k)|p\rangle  \langle 0|k-q\rangle.
\end{aligned}
\end{equation}
Note that we assume the direction of initial wave vector of the majority particle is the same with
the final one, i.e., $k-q$.
But for the case that the direction of initial wave vector of the majority particle is the same with the impurity
before scattering($p$),
and the scattering wave vector $q-k(\gg p)$ is mainly along the $x$-direction in the momentum space,
then the form factor $F_{\lambda\lambda'} (F'_{\lambda\lambda'})$ equals zero,
which means that for chiral system the pair propagator vanishes,
like in the case of helical system of $Bi_{2}Se_{3}$\cite{Shvonski A}.


In Fig.2-3 and Fig.4-5, we show the pair propagator and the corresponding self-energy, respectively,
for nonchiral and chiral cases.
By comparing the chiral case to the nonchiral one,
we can see that, when the angle $\theta$ (between the initial impurity wave vector and the scattering wave vector)
is nonzero,
both the pair propagator and self-energy diverge away from the nonchiral ones,
and such divergence locates mainly in the low-momentum region.
Thus the chiral effect leads to the instability for both the pair propagator and self-energy in small-$p$ region.
for low initial momentum of impurity.
In Fig.4-5, we can also see that, for stronger bare coupling $|g_{b}|$,
the polaron has a lower self-energy.
Since we consider the polaron formed in weak-coupling regime,
the final value of self-energy obtained here at large momentum agrees with 
the result of perturbation theory: $\Sigma(p,\omega)\propto g_{b}n$.
That is also in agreement with the mean-field result for the homogeneous condensate\cite{Rath S P}.
Note that during the calculation of pair propagator and self-energy, the direction of $p$ is unfixed, and thus
we donot integrate over the all possible angle $\theta$ (between $p$ and $q-k$),
as obviously can be seen from the figures.
The suppression of marginal Fermi liquid character can be seen from the self-energy as a function of $\omega$
as shown in lower-panel of Fig.5,
where the imaginary part of the self-energy diverges away from the intrinsic case $|{\rm Im}\Sigma|\sim \omega$\cite{Sarma S D}
at large value of $\omega$.
Through the imaginary part of the lower-panel of Fig.5,
we can also see that
the marginal Fermi liquid character is suppressed while the Fermi liquid character is arised with the increasing $g_{b}$.
The increase of $g_{b}$ also gives rise to the intraband single-particle excitation.

We note that, in Fermi-liquid picture with strong screening effect,
the Fermion interaction is dominated by the short-ranged one,
i.e., the Hubbard interaction,
then the strong spin fluctuation as well as the particle-hole fluctuation
are possible to build the bipolaron\cite{Camacho-Guardian A}.
The quasiparticle properties can also be detected by the spectral function,
which measures the probability of exciting or removing a (quasi)particle at a certain momentum.
Next we consider the particle spectral function containing the many-body effect,
\begin{equation} 
\begin{aligned}
A(p,\Omega)=-\frac{1}{\pi}\frac{|{\rm Im}\Sigma(p,\omega)|}
{(\Omega-{\rm Re}\Sigma(p,\omega)-\frac{\hbar^{2}p^{2}}{2m_{\phi}}+\mu_{\downarrow})^{2}+({\rm Im}\Sigma(p,\omega))^{2}}.
\end{aligned}
\end{equation}
From Fig.6-7, we can see that, except for zero frequency,
the spectral functions exhibit symmetry feature,
and the peaks decrease with the increase of $|g_{b}|$.
From the intensity plot of spectral function (Fig.8),
we can see that the spectral function
exhibits similar dispersion with the parabolic impurity (before scattering) in low-momentum region,
but exhibits linear dispersion in the large-$p$ region.
Also, we find that the width of spectral function becomes very narrow when close to zero-momentum.
Further,
by comparing to the dashed-line in the center inset of Fig.8,
we can easily see that the dispersion of original impurity ($\frac{p^{2}}{2m_{\downarrow}}-\mu_{\downarrow}$) is been lowered
by the polaronic effect (due to the negative self-energy correction),
and, in the mean time, the effective mass is also been increased due to the decrease of dispersion slope (compared to the original one).

At zero energy limit, the quasiparticle residue of the usual Dirac Fermions remains finite 
(here we assuming a noninteracting initial state)
and thus the coefficients $\psi_{0}$
and $\psi_{kq}$ remain finite too,
while for the multi-Weyl semimetal, the residue vanishes at zero-energy and then $\psi_{0}=\psi_{kq}=0$.
For zero momentum ($q=0$) with the lowest dispersion,
the impurity self-energy becomes,
\begin{equation} 
\begin{aligned}
\Sigma(p,\omega)=\frac{1}{S}\sum_{\mu_{\uparrow}}\frac{1}{\frac{\widetilde{m}}{2\pi\hbar^{2}a_{\psi\phi}}+\Pi(\omega+\Omega,p)}
\frac{1}{\omega+i0+\mu_{\uparrow}},
\end{aligned}
\end{equation}
with $\Pi(\omega+\Omega,p)=\Pi(\omega+\Omega,p,q=0)$, and 
$S$ is the volume of the space where all the chemical potential around the polaron are taken into account.
Now the center of mass is just $p$.
That is also agree with the results of the Fr\"{o}hlich polaron model\cite{Mogulkoc A2,Grusdt F2} in long-wavelength limit with strong electron-phonon coupling as well as the observable optical excitations\cite{Falck J P}.

For simplicity, we further set $\mu_{\uparrow}=0$ and the Fermi frequency $\Omega=1$ (which is possible in the zero-temperature limit),
then the polaron self-energy becomes
$\Sigma(p,\omega)=T(p,\omega)G^{0}_{\psi}(0,0)$ where $G^{0}_{\psi}(0,0)=1$
The self-energy is shown in Fig.7(a).
Then by substituting the above self-energy to the Eq.(22),
we obtain the corresponding particle spectral function as shown in the Fig.7(b),
where we set $\omega>0\ (\omega>\mu_{\uparrow})$ to make sure the spectral function here describes only the particle states.
Through Fig.7(b),
the total density of states can be obtained by integrating over the $p$-axis,
while the occupation probability\cite{Hassaneen K S A} can be obtained by integrating over the $\omega$-axis.

In Fig.9,
we show the effective masses induced by the polaronic effect and the quasiparticle residue,
which read
\begin{equation} 
\begin{aligned}
m^{*}/m=&\frac{1}{Z}(1+\frac{m}{p}\frac{\partial}{\partial p}{\rm Re}\Sigma(p,\omega))^{-1}\big|_{p=p_{F},\omega=0},\\
Z=&\frac{1}{1-{\rm Re}\partial_{\omega}\Sigma(p,\omega)}\bigg|_{\omega=E(p)},
\end{aligned}
\end{equation}
where $E(p)$ is determined self-consistently by the equation
\begin{equation} 
\begin{aligned}
E(p)=\frac{p^{2}}{2m_{\downarrow}}-\mu_{\downarrow}+{\rm Re}\Sigma(E(p),p),
\end{aligned}
\end{equation}
after the expansion coefficients of the polaron trial wave function are 
obtained by performing the variational minimization.
Note the above effective masses with the polaronic effect is equivalent to 
another form\cite{Cotle? O,Christensen R S,Sous}
\begin{equation} 
\begin{aligned}
m^{*}/m=\frac{1}{Z}(1+m\frac{\partial^{2}{\rm Re}\Sigma(p,\omega)}{\partial p^{2}})^{-1}\big|_{p=p_{F},\omega=0},
\end{aligned}
\end{equation}
as ${\rm Re}\Sigma(p,\omega)\sim p^{2}$.
The effective masses obtained by this formula is also showed in the Fig.9(a) 
by the dashed-line.
In Fig.9, we plot the induced effective mass and the quasiparticle residue as a function of momentum.
In small momentum region, 
both the induced effective mass and residue exhibit unusual behavior (due to the instabilities of the slow polaron with strong nonadiabatic 
dynamics).
In this region, the interaction is very strong and thus the residue could be very low,
and the induced effective mass may even become negative.
In the large momentum region,
the polaron becomes relatively stable,
then the power law behavior of $\Delta m^{*}$ and the logarithmic behavior of $Z$ can be seen.
We can also see that, for stronger attractive interaction (i.e., for larger $|g_{b}|$),
the polaron has a higher effective mass ($m^{*}=m_{\downarrow}+\Delta m^{*}$),
which is consistent with the result of Ref.\cite{Christensen R S,Scazza F},
and the residue asymptotically approaches one more slowly.
From the effective masses,
we can easily see that the positive interaction (repulsive polaron) induces stronger instability
compares to the negative one in the low-momentum region (especially for $\omega=0$).
In stable regime,
the induced effective masses $\Delta m^{*}$ increase more and more fast with the increasing initial impurity momentum $p$.
That is in agree with the result of Ref.\cite{Rath S P},
which, in addition, shows that the polaron is possible to changed to molecule if the $p$ keeps increasing.
In large-$p$ region, the many-body fluctuation is suppressed,
and the interaction effect is lowered (while the adiabatic effect is enhanced).
When the residue reaches one, the impurity will acts like a free particle.
Note that here the coupling strength is mainly controlled by the parameter $g_{b}$,
but also affected by the value of initial momentum $p$ as shown in the figure.
Further, at momentum $p>0.9$,
the curves of effective mass for non-chiral and chiral cases merge,
and the residue starts to approach one logarithmically,
which means that the chiral effect vanishes at large momentum (stable) regime.
For strong coupling with large enough $m^{*}$,
the polaron velocity due to the drag effect is totally depends on the velocity of majority particles\cite{Cotle? O}.

\section{Pair propagator and relaxation time at finite temperature}

For the case of finite temperature, we introduce
the Fermionic Matsubara frequencies as $\Omega=(2n+1)\pi T$, $\nu=(2n'+1)\pi T$
($n,n'$ are integer numbers),
which are discrete variables.
At finite temperature,
the $T$-matrix in resonance form can be generally written as\cite{Bastarrachea-Magnani M A,Wouters M}
\begin{equation} 
\begin{aligned}
T(p+q,\omega+\Omega)=\frac{1}{{\rm Re}\Pi_{va}(E_{b})-(\Pi(p+q,\omega+\Omega)-i\Gamma)},
\end{aligned}
\end{equation}
where the bare coupling is replaced by a vacuum pair propagator with zero center-of-mass momentum ($p+q=0$) $\Pi_{va}(E_{b})$
which reads
\begin{equation} 
\begin{aligned}
\Pi_{va}(E_{b}|_{p,q\rightarrow 0})=&
\int\frac{d^{3}k}{(2\pi)^{3}}
\frac{-1}{E_{b}+\varepsilon_{-k\downarrow}+\varepsilon_{k\uparrow}}\\
=&-4\pi 
m_{\uparrow}[\Lambda-\sqrt{E_{b}m_{\uparrow}}{\rm arctan}\frac{\Lambda}{\sqrt{E_{b}m_{\uparrow}}}],
\end{aligned}
\end{equation}
which is the same with the expression of bare coupling in attractive case ($g^{-1}_{b}<0$)
at zero majority particle density (two-body scattering instead of many-body scatterig).
$E_{b}$ is the binding energy which has $E_{b}=-{\rm Re}\Sigma(p=0,\omega)\sim |g|\gg 1$
for attractive polaron at unitarity,
and $E_{b}\gg |\mu_{\downarrow}|=|g|n_{\downarrow}$ in weak coupling regime.
Here the renormalized static chemical potential is related to the mass enhancement and the population
of the impurities, as well as the grand canonical potential\cite{scripta}.
While for the repulsive polaron, it has $E_{b}=\frac{1}{m_{\downarrow}a^{2}}\sim g^{-2}$.
The decay rate $\Gamma={\rm Im}\Sigma=1/\tau_{p}=1/2\tau_{q}$ ($\tau_{q}$ is the quasiparticle lifetime)
here contains the temperature effect, like the Landau damping\cite{Ozeri R},
as well as the other disorders.
We note that, at zero temperature limit, the decay is mainly through the way that an
 excitation dissociated into two or more excitations with lower energy,
e.g., the polaron decays back to the bare impurity and an electron-hole pair.
This is similar to the so called Beliaev damping process,
which is usually happen in the ultracold Bosonic systems\cite{Katz N}.
While at finite temperature, since there is an initial occupation for the thermally transferred momentum, the scattering (collision) cross section of the Landau damping is nonzero.
For the case of weak-couping ($g\ll\hbar^{2}/m^{*}$),
the polaron energy at $E=-E_{b}$ is a pole of the low-energy scattering $T$-matrix,
and the chemical potential of the parabolic impurity is much lower than the binding energy.
It shows that the mean-field theories are valid in this picture,
where the polaron shift is much smaller than its kinetic energy (or the hopping amplitude integral $J$):
$\frac{g^{2}}{2m_{\downarrow}\omega^{2}}\ll J$.

The pair propagator can then be written as
\begin{equation} 
\begin{aligned}
\Pi(p+q,\omega+\Omega)
=\int\frac{d^{3}k}{(2\pi)^{3}}[\sum_{n'}\frac{T}{V}
                             G_{0}^{\psi}(\nu,k)G_{0}^{\phi}(\omega+\Omega-\nu,p+q-k)],
\end{aligned}
\end{equation}
where we consider the single band model and regard the Green's functions as the only eigenvalue of the matrix.
Here we define 
\begin{equation} 
\begin{aligned}
G_{0}^{\psi}(\nu,k)=&\frac{1}{i\nu-\frac{k^{2}}{2m_{\psi}}+\mu_{\uparrow}},\\
G_{0}^{\phi}(\omega+\Omega-\nu,p+q-k)=&\frac{1}{i\omega+i\Omega-i\nu-\frac{(p+q-k)^{2}}{2m_{\phi}}+\mu_{\downarrow}}.
\end{aligned}
\end{equation}
The summation over Matsubara frequencies ($i\nu$) can be calculated as
\begin{equation} 
\begin{aligned}
\sum_{\nu}G_{0}^{\psi}(\nu,k)G_{0}^{\phi}(\omega+\Omega-\nu,p+q-k)=
\sum_{n'=-\infty}^{\infty}\frac{1}{i(2n'+1)\pi T-a}\frac{1}{-i(2n'+1)\pi T-b}=
\frac{{\rm tanh}\frac{b}{2T}+{\rm tanh}\frac{a}{2T}}
{2T(a+b)},
\end{aligned}
\end{equation}
where $a\equiv \varepsilon_{k\uparrow}=\frac{k^{2}}{2m_{\psi}}-\mu_{\uparrow}$,
  $b\equiv -i\omega-i\Omega+\frac{(p+q-k)^{2}}{2m_{\phi}}-\mu_{\downarrow}$,
and the above result can also be rewritten as
\begin{equation} 
\begin{aligned}
\sum_{\nu}G_{0}^{\psi}(\nu,k)G_{0}^{\phi}(\omega+\Omega-\nu,p+q-k)=
\frac{e^{\frac{a+b}{T}}-1}{T(a+b)(e^{a/T}+1)(e^{b/T}+1)}=
\frac{N_{F}(a)N_{F}(b)}{N_{B}(a+b)T(a+b)},
\end{aligned}
\end{equation}
where $N_{F}(x)=1/(e^{x/T}-1)$ is the Bose distribution function
and $N_{F}(x)=1/(e^{x/T}+1)$ is the Fermi distribution function.
For small $a$ and $b$, i.e., in the limit of small energy and small frequency,
we approximate ${\rm tanh}(x)\approx x-\frac{x^{3}}{3}+O(x^{5})$,
then the pair propagator (nonchiral) becomes
\begin{equation} 
\begin{aligned}
%
\Pi(p+q,\omega+\Omega)
=&4\pi\frac{1}{(2\pi)^{3}}\int^{\Lambda}_{k_{F}} k^{2}dk
\left[\frac{{\rm tanh}\frac{b}{2T}+{\rm tanh}\frac{a}{2T}}
{2T(a+b)}\right]\\
=&4\pi\frac{1}{(2\pi)^{3}}\int^{\Lambda}_{k_{F}} k^{2}dk
\left[\frac{\frac{b}{2T}-\frac{1}{3}(\frac{b}{2T})^{3}+\frac{a}{2T}-\frac{1}{3}(\frac{a}{2T})^{3}}
{2T(a+b)}\right]\\
=&4\pi\frac{1}{(2\pi)^{3}}\int^{\Lambda}_{k_{F}} k^{2}dk
\left[\frac{1}{4T^{2}}-\frac{1}{3}\frac{b^{3}+a^{3}}{(2T)^{4}}\frac{1}{a+b}
\right]\\
=&4\pi\frac{1}{(2\pi)^{3}}\int^{\Lambda}_{k_{F}} k^{2}dk
\left[\frac{1}{4T^{2}}-\frac{1}{3}\left(\frac{1}{2T}\right)^{4}(a^{2}-ab+b^{2})
\right]\\
=&\frac{1}{8\pi^{2}T^{2}}\frac{(\Lambda-k_{F})^{3}}{3}
-\mathcal{F},
\end{aligned}
\end{equation}
where
\begin{equation} 
\begin{aligned}
\mathcal{F}=&\frac{1}{6\pi^{2}}\frac{1}{(2T)^{4}}
             \int^{\Lambda}_{k_{F}}k^{2}(a^{2}-ab+b^{2})dk\\
=&\frac{1}{6\pi^{2}}\frac{1}{(2T)^{4}}
\frac{k^{3}}{420m_{\psi}^{2}m_{\phi}^{2}}
[
35m^{2}_{\psi}(4c^{2}m_{\phi}^{2}+2cm_{\phi}((p+q)^{2}-2dm_{\phi})\\
&+((p+q)^{2}-2dm_{\phi})^{2})
+21k^{2}m_{\psi}(2cm_{\phi}(m_{\psi}-2m_{\phi})+2dm_{\phi}(m_{\phi}-2m_{\psi})+(6m_{\psi}-m_{\phi})(p+q)^{2})\\
&-105km_{\psi}^{2}(p+q)(cm_{\phi}-2dm_{\phi}+(p+q)^{2})
+15k^{4}(m_{\psi}^{2}-m_{\phi}m_{\psi}+m_{\phi}^{2})\\
&-35k^{3}m_{\psi}(2m_{\psi}-m_{\phi})(p+q)
]\bigg|^{\Lambda}_{k_{F}}.
\end{aligned}
\end{equation}
where we define $c\equiv \mu_{\uparrow}$, $d\equiv i\omega+i\Omega+\mu_{\downarrow}$.
As shown in the Fig.10,
the momentum- or energy-dependence of the polarization function decreases with the increase of temperature.
At high-enough temperature, both the imaginary and real part of the polarization function become constant,
and thus we can suspect that 
the induced effective mass will become infinite (self-trapped polaron) at high enough temperature
while the residue will equals to one in the same case.

The ladder approximation (by summing over the ladder diagrams to higher order
which correspond to the forward scattering) 
results in accurate results of the pair propagator and
self-energy,
and it also agrees with the Quantum Monte-Carlo calculation as well as
the experimental results.
The single-channel $T$-matrix (which contains the pair propagator) introduces the tuneable $s$-wave scattering length to the 
manipulation of the behavior of a single impurity embedded to a Fermi sea,
which describes the scattering between a pair of atoms with up and down spins, respectively,
and within the center of mass frame with energy $\varepsilon=\omega-(p+q)/2(m_{\psi}+m_{\phi})+\mu_{\uparrow}+\mu_{\downarrow}$.
At finite temperature and for the case that the number density of the bose impurity 
is much lower than the Fermions (without the effect of three-atom loss (the Efimov trimers)\cite{Fratini E,Massignan P,Pietil? V}), the pair 
propagator can also be written as
\cite{Massignan P2,Fratini E}
\begin{equation} 
\begin{aligned}
\Pi(p+q,\omega+\Omega)=
\sum_{k}\frac{1-N_{F}(\varepsilon_{k\uparrow})-N_{F}(\varepsilon_{p+q-k\downarrow})}
{\omega+i0-\varepsilon_{p+q-k\downarrow}-\varepsilon_{k\uparrow}+\varepsilon_{q\uparrow}},
\end{aligned}
\end{equation}
where $p$ and $q$ correspond to the momentum of impurity and hole respectively,
$\omega$ and $\Omega$ correspond to the frequency of impurity and hole respectively.
For pairing mechanism, this expression is definitely important,
e.g., for the pairing instability\cite{Cui X,Pietil? V,Adachi K,Pekker D} and the resonantly enhanced correlation,
and its real part and imaginary part are easy to obtained
by firstly replacing the imaginary frequencies in denominator with the 
analytical continuation and then using the Dirac identity (for retarded functions)
 $\lim_{\eta\rightarrow 0}\frac{1}{x\pm i\eta}=P(\frac{1}{x})\mp i\pi\delta(x)$.
We can see that the factor $F_{\lambda\lambda'}$ is in fact related to the angle between the wave vectors of
polaron (coherently dressed by the particle-hole excitations
of majority part) and the electron with momentum $k$.
And this term is unnecessary in the three (or two)-dimensional electron (or hole) gases,
it
is nonzero only when the eigenstates at different wave vectors have overlap
(corresponds to the two statistical functions in the numerator),
which for the gapped systems (or consider the longitudinal wave vector $k_{z}$) reads
\begin{equation}
\begin{aligned}
|p\rangle=
&\begin{pmatrix}
{\rm cos}\frac{\theta}{2}\\
\lambda{\rm sin}\frac{\theta}{2}e^{2\lambda i\phi_{p}}
\end{pmatrix}\\
=&\begin{pmatrix}
{\rm cos}\frac{\theta}{2}e^{-\lambda i\phi_{p}}\\
\lambda{\rm sin}\frac{\theta}{2}e^{\lambda i\phi_{p}}
\end{pmatrix},
\end{aligned}
\end{equation}
where the indices $\pm$ denote the sign of band energy (i.e., the conduction band and valence band),
and $\phi={\rm atan}k_{y}/k_{x}$,
    $\theta={\rm atan}\sqrt{k^{2}_{x}+k^{2}_{y}}/k_{z}$.
Then the overlap factor reads
$F_{\lambda\lambda'}=\frac{1\pm ({\rm cos}\theta{\rm cos}\theta'-{\rm sin}\theta{\rm sin}\theta'{\rm sin}\ b)}{2}$
where $\theta'={\rm atan}\sqrt{k'^{2}_{x}+k'^{2}_{y}}/k'_{z}$.
That is clearly different from the ones appear in two-dimensional system\cite{Culcer D}.
For the calculation in main text, we use the two-dimensional chiral factor $F_{\lambda\lambda'}=\frac{1\pm {\rm cos}\ b}{2}=\frac{1}{2}(1\pm \frac{k+q{\rm cos}\ a}{\sqrt{k^{2}+q^{2}+2kq{\rm cos}\ a}})$
due to the nature of weak-chirality of the system we discussed.
For another case of three-dimensional system, at long-wavelength limit ($k_{z}\rightarrow 0$) and with isotropic dispersion, 
such approximation is also applicable as shown in, e.g., Ref.\cite{Ahn S}.
When the vertex correction is not taken into account,
for inversed-frequency of $\Pi(p+q,-\omega-\Omega)$,
we can use the identity $\Pi(p+q,-\omega-\Omega)=\Pi^{*}(p+q,\omega+\Omega)$,
i.e., ${\rm Re}\ \Pi(p+q,-\omega-\Omega)={\rm Re} \Pi(p+q,\omega+\Omega)$,
${\rm Im}\ \Pi(p+q,-\omega-\Omega)=-{\rm Im} \Pi(p+q,\omega+\Omega)$
Further, when the chirality (from the Weyl system) appears, the causality relations are studied in Ref.\cite{Zhou J}.

Then the self-energy at finite temperature can be obtained as
\begin{equation} 
\begin{aligned}
\Sigma(p,\omega)=\frac{T}{V}\int^{k_{F}}_{0}\frac{d^{3}q}{(2\pi)^{3}}\sum_{n}
T(p+q,\omega+\Omega)G^{0}_{\psi}(q,\Omega),
\end{aligned}
\end{equation}
where $G^{0}_{\psi}(q,\Omega)$ can also be replaced by $G_{\psi}(q,\Omega)$
which contains the self-energy term when consider the self-energy effect as done in Ref.\cite{Enss T} with strong scattering strength.
In addition, we discuss the case when consider the ladder vertex correction,
where summation over Matsubara frequency can be done by using the method of contour integral (in optical limit)\cite{Mahan G D},
\begin{equation} 
\begin{aligned}
\frac{T}{V}\sum_{n'}G^{\psi}(\nu)G^{\phi}(\omega+\Omega-\nu)\Gamma(\nu,\omega+\Omega-\nu)=-\oint_{\mathcal{C}}\frac{dz}{2\pi i}G^{\psi}(z)G^{\phi}(\omega+\Omega-z)\Gamma(z,\omega+\Omega-z),
\end{aligned}
\end{equation}
where $\Gamma(\nu,\omega+\Omega-\nu)$ denotes the vertex function.

At finite temperature where the $s$-wave scattering is still dominating,
the low-energy excitations induced by quantum fluctuation has a more significant effect on the properties of polaron 
compared to the thermal excitations
especially for the case of small-chemical potential,
like the particle-hole parts (especially at low dimension\cite{Koschorreck M,Endres M}) or the phonon-like (Fr{\o}hlich type) excitations.
For chiral system at finite temperature, the transport relaxation time (in ladder diagram) of impurity 
due to the scattering by the electron-hole pair 
contains a $(1-{\rm cos}\theta)$ term, which suppresses the forward scattering ($\theta=0$)
and exists as long as the elastic scattering is involved in the scattering event.
This term cannot be found in the quasiparticle relaxation time in single bubble diagram,
and it together with the chiral term determines the scattering cross section\cite{Adam S}.
While for the gapless Dirac system, like the intrinsic graphene, both the forward and backward scattering are suppressed\cite{Adam S,Hwang E H2,Iurov A}.
The Boltzmann transport theory gives the following inverse relaxation time (in second order Born approximation)
\begin{equation} 
\begin{aligned}
\frac{1}{\tau_{p}}=\frac{2\pi}{\hbar}\sum_{k,q}(1-{\rm cos}\theta)W_{p+q-k,p},
\end{aligned}
\end{equation}
where $p'=p+q-k$ is the wave vector of impurity after scattering, and 
$\theta=\phi_{p+q-k}-\phi_{p}$ is the angle between wave vectors before and after
scattering.
\begin{equation} 
\begin{aligned}
W_{p+q-k,p}=&(1-N_{F}(\varepsilon_{k-q\uparrow}))\delta(\omega-\varepsilon_{p+q-k\downarrow}-\varepsilon_{k\uparrow}+\varepsilon_{q\uparrow})
|g_{b}(k-q)|^{2}|\langle p+q-k|p\rangle|^{2}
\end{aligned}
\end{equation}
 is the transition rate and can be approximated as the imaginary part of self-energy ${\rm Im}\Sigma(p,\omega)$
and the form factor can be found in Eq.(16).
Then the relaxation time can be obtained as
\begin{equation} 
\begin{aligned}
\frac{1}{\tau_{p}}=&\frac{2\pi}{\hbar}\sum_{k,q}(1-{\rm cos}(\phi_{p+q-k}-\phi_{p}))
(1-N_{F}(\varepsilon_{k-q\uparrow}))\\
                   &\delta(\omega-\varepsilon_{p+q-k\downarrow}-\varepsilon_{k\uparrow}+\varepsilon_{q\uparrow})
|g_{b}(k-q)|^{2}{\rm cos}^{2}(\phi_{p+q-k}-\phi_{p})\\
\approx &
\frac{2\pi}{(2\pi)^{3}}\int^{\Lambda}_{k_{F}} k^{2}dk\int^{\pi}_{0}{\rm sin}\Phi d\Phi\int^{2\pi}_{0}d\phi_{p+q-k}
(1-N_{F}(\varepsilon_{k-q\uparrow}))\\
                  &\delta(\omega-\varepsilon_{p+q-k\downarrow}-\varepsilon_{k\uparrow}+\varepsilon_{q\uparrow})
|g_{b}(k-q)|^{2}{\rm cos}^{2}(\phi_{p+q-k}-\phi_{p})\\
=&\frac{4\pi}{(2\pi)^{3}}\int^{\Lambda}_{k_{F}} k^{2}dk \int^{2\pi}_{0}d\phi_{p+q-k}
(1-N_{F}(\varepsilon_{k-q\uparrow}))\\
                   &\delta(\omega-\varepsilon_{p+q-k\downarrow}-\varepsilon_{k\uparrow}+\varepsilon_{q\uparrow})
|g_{b}(k-q)|^{2}{\rm cos}^{2}(\phi_{p+q-k}-\phi_{p}),
\end{aligned}
\end{equation}
where we assume the case of static hole ($q=0$) for simplicity.
Also, we assume the scattering wave vector is larger than the initial impurity wave vector ($k>p$) so that
the integral of $\phi_{p+q-k}$ can goes over the whole range.
$g_{b}(k-q)$ is the scattering wave vector-dependent bare interaction vertex (kernel).
When the chiral effect is considered, the interaction vertex should be replaced by the one 
presented in Eq.(58).

Note that in case of electron-phonon interaction, where both the electron-phonon polaron and the electron-ion polaron are produced
with the emergent electronic screening and lattice screening\cite{Sio W H},
the interaction vertex here depends on both the scattering wave vector and the initial wave vector (like $p$)\cite{Marchand D J J}.
While for the case of multi-impurity (Eq.(22)),
the above relaxation time should be rewritten as
\begin{equation} 
\begin{aligned}
\frac{1}{\tau_{p}}=\frac{2\pi}{\hbar}\sum_{q,k}(1-{\rm cos}\theta)W_{p+q-k,p}\left(\frac{1-N_{F}(\varepsilon_{p+q-k})}{1-N_{F}(\varepsilon_{p})}\right),
\end{aligned}
\end{equation}
which can be obtained through the following relation in large-impurity momentum regime,
\begin{equation} 
\begin{aligned}
1-N_{F}(\varepsilon_{k\uparrow})-N_{F}(\varepsilon_{p+q-k\downarrow})=(1-N_{F}(\varepsilon_{k-q\uparrow}))
\left(\frac{1-N_{F}(\varepsilon_{p+q-k})}{1-N_{F}(\varepsilon_{p})}\right).
\end{aligned}
\end{equation}
Also here the existence of factor $\left(\frac{1-N_{F}(\varepsilon_{p+q-k})}{1-N_{F}(\varepsilon_{p})}\right)$ implies
thatthe scattering is inelastic
\cite{Kim S,Gunst T,Hwang E H}
in multi-impurity case.
The impurity scattering angle $\theta$ here is defined as the angle between $p$ and $p+q-k$,
which can be written as $\theta={\rm arccos}({\bf n}_{p+q-k}/{\bf n}_{p})$ where ${\bf n}_{p}$ is the direction projection,
and it is related to the scattering wave vector by $q-k=2p_{F}{\rm sin}(\theta/2)$.
Note that for strong coupling case the temperature-dependence of the elastic scattering rate
is suppressed.
That is because the coupling strength is inversely proportional to the impurity frequency $\omega$ and hopping integral.
In this case the coupling becomes relative momentum-dependent\cite{Petrov D S}.
Note that even in equilibrium scenario with $\tau_{p}\rightarrow\infty$,
the polaron (with small density of impurity) can still be affected by the drag effect
from the drived Fermi bath\cite{Cotle? O} or the excited bosons\cite{Nakano} like
 the phonons at finite temperature.
In the low-doping case with non-Fermi liquid feature, 
the transport relaxation rate is bound by the Planckian limit 
$1/\tau \lesssim T$.

Considering the phonon effect (inelastic scattering),
the above relaxation time can be rewritten as
 \begin{equation} 
\begin{aligned}
\frac{1}{\tau_{p}}=\frac{2\pi}{\hbar}\sum_{p'}
(1-{\rm cos}\theta\frac{\tau_{p'}}{\tau_{p}}\frac{N_{F}(\varepsilon_{p})}{N_{F}(\varepsilon_{p'})}
\frac{1-N_{F}(\varepsilon_{p})}{1-N_{F}(\varepsilon_{p'})})
W_{pp'}\left(\frac{1-N_{F}(\varepsilon_{p'})}{1-N_{F}(\varepsilon_{p})}\right),
\end{aligned}
\end{equation}
where $p'=p-q'$.
Here 
$\frac{\tau_{p'}}{\tau_{p}}\frac{N_{F}(\varepsilon_{p})}{N_{F}(\varepsilon_{p'})}
\frac{1-N_{F}(\varepsilon_{p})}{1-N_{F}(\varepsilon_{p'})}=1$ and 
${\rm cos}\theta=\frac{{\bf v}_{p'}\cdot{\bf v}_{p}}{|{\bf v}_{p'}||{\bf v}_{p}|}
=\frac{\tau_{p'}|{\bf v}_{p'}|}{\tau_{p}|{\bf v}_{p}|}$.
Further discussion about the phonon and the thermal excitation are presented in the Appendix.C.

\section{Summary}


In this paper, we discuss the polaron formed in a doped chiral/nonchiral parabolic system.
We discuss in detail the diagrammatic approach for the Fermion polaron self-energy
within extended ladder approximation (to arbitrary order of bare coupling $g_{b}$).
 But we emphasize that it is different from
the perturbation theory as shown in Ref.\cite{Christensen R S} although we use the 
variational approach and focus on the weak coupling regime
throughout this paper,
and the non-self-consistent $T$-matrix approach is also
based on the perturbative expansion\cite{Rath S P} of the impurity Green's function.
A strong evidence is the real part of self-energy as shown in Fig.4 and Fig.5,
where the resonance structure of the spectrum (with two quasiparticle solutions)
can be seen around the pole of the $T$-matrix as discussed in Sec.5 and Appendix.B.
Note that, however, in a Bose-Einstein condensate
with weak coupling,
the energy of Bose polaron calculated by the variational approach may in consistent
with that obtained by perturbation theory\cite{Field B}.
The method reported here can also be applied to the Dirac systems
with finite effective mass.
In the numerical simulations,
we studied the effect of bare coupling as well as the instability of the pair propgator and spectral function
with a small-renormalized effective mass and chemical potential.
The many-body effect is also analyzed through the study of spectral function.
Although the $T$-matrix approximation here takes into account the pairing interaction 
with the leading instability even in the presence of weak intraspecies interactions (the $p$-wave interaction), 
it is indeed a nonperturbative theory (see Eq.(54)) which is evidenced by the absence of the self-consistency
(i.e., the Coulomb induced exchange self-energy is the Hartree-Fock type and in lack of the dynamical dielectric function),
thus the energy is nonconserving 
and the quasiparticle weigh is lower than the one in random phase approximation (RPA) theory 
(with dynamical screening) or the partial self-consistent theory (with static screened interaction or dielectric function)\cite{Holm B}.
The formation and properties of the attractive polaron formed in a two-dimensional semi-Dirac system is reported in Ref.
\cite{polaron2, polaron3}, 
where the anisotropic effective masses distribution takes an important role,
and we approximate the dispersion as the parabolic anisotropic one 
similar to the plasmon-polaron formed by the phosphorene locates on polar substrates\cite{Saberi-Pouya S}.
Besides, 
the $p$-wave scattering of the polaron system is also been studied in topological superfluid and the weak-coupled BEC recently\cite{Kinnunen J J,Qin F}.
The attractive polaron as a quasiparticle can be observed experimentally through the momentum-resolved photoemission spectroscopy\cite{Koschorreck M}.
For Bose-Hubbard model in superfluid phase, 
the Bose polaron with a spin impurity can be created by the off-resonance laser and microscope objective\cite{Fukuhara T,Weitenberg C}.
At half-filling ($\mu=0$) where the electron density equals 1 and the on-site Hubbard U is much larger than the mobility of impurity, the 
spin impurity is localized and in this case the non-self-consistent $T$-matrix approximation has high accuracy 
due to the weak dynamical screening effect from the carriers.
In one-dimensional geometry, 
this experiment also provides a platform to explore the other polaronic physics like the propagation velocity affected by the self-trapping effect,
which implies that the polaronic effect can emergent also in the superconductors or the Mott insulators\cite{Endres M}.
The self-trapping effect will becomes more obvious at finite temperature due to the emergent electron-phonon coupling\cite{polaron3}.
While at low-temperature limit (e.g., $<1\mu K$)
the magnetic or electric trapping can be applied to the molecule cloud or the hyperfine states 
of the alkali atoms,
to design the quantum memory setups in quantum circuit\cite{Rabl P}.
For solid state like the Dirac system, in the presence of, 
e.g., the  separable $s$-wave potential\cite{Gaul C}, 
the $s$-wave scattering as well as the elastic scattering can be treated as dominating at low-temperature limit, 
and the two-body Lippmann-Schwinger equation is still valid in obtaining the coupling parameters and the $T$-matrix.
In fact for impurity and the particle-hole part (excited by the quantum fluctuation) with energies similar to the same (gapped) Dirac cone, 
the scattering can be treated as the intravalley one, 
which can help us to deal with the multichannel problem.

\section{Appendix.A: Variational approach in mean-field approximation for isotropic lattice}

Energy and $g_{b}$ are the variational parameters during the minimization procedure 
(carry the first or second derivatives) of the expectation value of the
Hamiltonian matrix element, or free energy, to arrives the ground state\cite{scripta}.

For Bose polaron in BEC,
the method of mean-field approximation is valid in the presence of the weak on-site Boson-Boson or Boson-Fermion (impurity)
interaction, i.e., the dilute BEC,
and certainly,
the physical parameters like the lattice parameter or the interaction strength
can be controlled by the Feshbach technique,
and the strong-interaction regime can also reached by this method.
Here we use the variational approach base on Gaussian variational Ansatz and the Lagrangian optimization.
The variational approach can be generalized by the differential of the matrix element\cite{Li W}
\begin{equation} 
\begin{aligned}
\frac{\partial \langle \Psi|H-E|\Psi\rangle}{\partial (i\psi^{*})}=0,
\end{aligned}
\end{equation}
where $\Psi$ is the trial wave function including the interaction effect,
$H$ is the effective Hamiltonian of the discussing system,
$E$ is the Lagrange multiplier which gives the local minimal energy,
$\psi$ is the real components.
We make the mean-field approximation to the Grassmann field which written as $c_{j}$ at site $j$,
then the Lagrangian reads
\begin{equation} 
\begin{aligned}
L=\sum_{j}i\frac{\partial H^{MF}}{\partial (ic_{j}^{*})}c_{j}^{*}-H^{MF},
\end{aligned}
\end{equation}
where $H^{MF}$ is the mean-field Hamiltonian,
the Grassmann field is treated as a dynamical Gaussian profile as
\begin{equation} 
\begin{aligned}
c_{j}=\frac{\sqrt{2}}{r\sqrt{\pi}}{\rm exp}[-\frac{(j-c)^{2}}{r^{2}}+ik(j-c)],
\end{aligned}
\end{equation}
where $c$ and $k$ are the coordinate and momentum of the center of wave package, respectively,
$r$ is the width of wave package.

Then base on the Euler-Lagrangian relation 
\begin{equation} 
\begin{aligned}
\frac{\partial L}{\partial c}-\frac{d}{dt}\frac{\partial L}{\partial \dot{c}}=-m\ddot{c}=0,
\end{aligned}
\end{equation}
we obtain 
\begin{equation} 
\begin{aligned}
c={\rm sin}k\ e^{-\frac{1}{2r^{2}}}t,\\
m=\frac{1}{\hbar^{2}}{\rm cos}k\ e^{-\frac{1}{2r^{2}}},
\end{aligned}
\end{equation}
In the absence of the external potential,
the effective Hamiltonian is independent of $c$,
\begin{equation} 
\begin{aligned}
\frac{\partial H^{MF}}{\partial c}=&0,\\
\frac{\partial H^{MF}}{\partial \dot{c}}=&\left[-\frac{A}{r^{2}}-{\rm cos}k\ e^{\frac{-1}{2r^{2}}}\frac{1}{r^{3}}\right]
\frac{r^{3}}{{\rm sin}k\ e^{\frac{-1}{2r^{2}}}},
\end{aligned}
\end{equation}
where the effective coupling parameter $A=U/(4J\sqrt{\pi})$ as a ratio between the on-site interaction $U$ and the tunneling strength $J$,
the mean-field Hamiltonian here reads $H^{MF}=\frac{A}{r}-{\rm cos}e^{\frac{-1}{2r^{2}}}$.
Here we note that the critical value of effective coupling parameter $A$ for the 
self-trapping, soliton, and breather are not continued during the BEC-BCS crossover,
 unlike the attractive self-energy beyong the Hartree-Fock approximation\cite{Li W}.
In the strong interacting case,
the electron may become self-trapped and with localized wave package
characterized by a diverging effective mass $m$.

For lattice model with finite sites in mean-field description 
($\sum_{p}\varepsilon_{p}c_{p}^{\dag}c_{p}=-J\sum_{ij}c^{\dag}_{i}c_{j}$),
we can carry a simplest variational procedure, which may shows some differences compared to the one shown
in Appendix.C.
For the intrinsic lattice system and the dressed one (by the polaronic dynamics),
the low-energy Hamiltonian read
\begin{equation} 
\begin{aligned}
H=-J\sum_{ij\sigma}c_{i\sigma}^{\dag}c_{j\sigma}+g_{b}\sum_{i}n_{i\uparrow}n_{i\downarrow},\\
H'=-J'\sum_{ij\sigma}c_{i\sigma}^{\dag}c_{j\sigma}+g'_{b}\sum_{i}n_{i\uparrow}n_{i\downarrow},\\
\end{aligned}
\end{equation}
where $J$ and $J'$ are the intrinsic and modified hopping.
Then we have
\begin{equation} 
\begin{aligned}
\frac{1}{N}(H'-H)=&-(J'-J)\sum_{ij\sigma}\langle c_{i\sigma}^{\dag}c_{j\sigma}\rangle
+(g'_{b}-g_{b})\sum_{i}\langle n_{i}n_{j}\rangle,\\
\frac{1}{N}\Sigma=&\frac{J'}{N}\sum_{ij\sigma}\langle c_{i\sigma}^{\dag}c_{j\sigma}\rangle,\\
\frac{1}{N}\partial_{J'}(H'-H)
=&J\partial_{J'}\sum_{ij\sigma}\langle c_{i\sigma}^{\dag}c_{j\sigma}\rangle
 -\sum_{ij\sigma}\langle c_{i\sigma}^{\dag}c_{j\sigma}\rangle-J'\partial_{J'}\sum_{ij\sigma}\langle c_{i\sigma}^{\dag}c_{j\sigma}\rangle
 +(g'_{b}-g_{b})\partial_{J'}\sum_{i}\langle n_{i}n_{j}\rangle,\\
\frac{1}{N}\partial_{J'}\Sigma
=&\frac{1}{N}\sum_{ij\sigma}\langle c_{i\sigma}^{\dag}c_{j\sigma}\rangle,
\end{aligned}
\end{equation}
where $\sigma$ here is the interaction energy correction.
By solving $\frac{1}{N}(H'-H+\sigma)=0$ we obtain the optimal value 
of $J'$ which is
\begin{equation} 
\begin{aligned}
J'=(g'_{b}-g_{b})\frac{\partial_{J'}\langle n_{i\uparrow}n_{i\downarrow}\rangle}{\partial_{J'}\langle n_{i\uparrow}n_{j\uparrow}+n_{i\downarrow}n_{j\downarrow}\rangle}.
\end{aligned}
\end{equation}

\section{Appendix.B: Self-energy correction in Ladder diagram}

As stated above, the mean-field description (which is restricted to local interactions)
is valid for sufficiently weak interaction strength,
in which case the self-consistent chemical potential is much smaller than the binding energy 
(the real pole of the zero-momentum low energy scattering matrix or $T$ matrix).
The self-consistent chemical potential here is renormalized by the simplest Beliaev type 
anomalous self-energy to the first approximation,
$\mu_{\downarrow}=\Sigma_{a}=n_{\downarrow}g_{b}=n_{\downarrow}\langle 0|T|0\rangle$,
which is also the lowest contribution from the interaction to the chemical potential,
or the so-called vacuum scattering matrix.
Extending to the strong couping regime, although the mean-field description is fail,
the above relation is still valid as we replace the bare coupling within $T$-matrix by
the binding energy with 
$\omega=E_{b},\ p=0$\cite{biexcitons}.
The self-consistent chemical potential 
also related to the noninteracting part of the grand canonical potential $\Omega_{c}$
logarithmically through the Green's function.
Since $n_{\downarrow}=-\partial\Omega_{c}/\partial \mu_{\downarrow}$,
we have $\Omega_{c}\sim -\frac{1}{g_{b}}\frac{\mu_{\downarrow}^{2}}{2}$.
Note that for the case of large momentum transfer,
the static part of the particle-hole bubble (within RPA) can not be ignored
when the local exchange-correlation potential dominates over the long-range Coulomb potential
(Hartree potential)
and the constant potential $g_{b}$.

While in the second-order of the diagram expansion,
we have
\begin{equation} 
\begin{aligned}
\Sigma(p,\omega)=&\frac{n}{N}\int\frac{d^{3}q}{(3\pi)^{3}}\frac{d\Omega}{2\pi}T(p+q,\omega+\Omega)G(q,\Omega)\\
=&\frac{n}{N}\int\frac{d^{3}q}{(3\pi)^{3}}\frac{d\Omega}{2\pi}
                    \sum_{k}g_{k}g_{p+q-k}G(k,\nu)G(p+q-k,\omega+\Omega-\nu)
G(q,\Omega)\\
=&\frac{n}{N}\int\frac{d^{3}q}{(3\pi)^{3}}\frac{d\Omega}{2\pi}
\frac{1}{g_{b}^{-1}-\Pi(p+q,\omega+\Omega)}.
\end{aligned}
\end{equation}
In the last line of the above equation we use the approximation of bare coupling (constant),
which has a similar form with the bare susceptibility in RPA.
Note that such approximation breaks the self-consistency of a many-body system\cite{Katanin A A}.
While in the second line, the $T$-matrix is obviously been written as a
 renormalized interaction vertex (to lowest order in $1/N$; $N$ is the flavor number),
i.e., the second term of the $T$-matrix.
This can also be applied to the non-Fermi-liquid metal (incoherent metal)\cite{Chowdhury D}.

Compared to the above case (first-order),
the factor of $N$ is brought by the particle-particle loops (i.e., a pair
of impurity-majority propagators) in the second-order contribution,
i.e., summing up all the flavors,
while the vertice brings the factor of $1/N$, which is absent in conventional RPA.
The vertex correction to the bare correlation function 
(correlator $\Pi(p+q,\omega+\Omega)$) within non-self-consistent $T$-matrix
approximation can be written as
\begin{equation} 
\begin{aligned}
T(p+q,\omega+\Omega)=&
g_{b}+
\sum_{k}g_{k}g_{p+q-k}G(k,\Omega)G(p+q-k,\omega+\Omega)\\
&+\sum_{k,k'}g_{k}g_{p+q-k}G(k,\Omega)G(p+q-k,\omega+\Omega)
g(k')G(k',\Omega)G(p+q-k',\omega+\Omega)+\cdot\cdot\cdot\\
=&
g_{b}+
\sum_{k}g_{k}g_{p+q-k}G(k,\Omega)G(p+q-k,\omega+\Omega)\\
&+\sum_{k,k'}g_{k}g_{p+q-k}G(k,\Omega)G(p+q-k,\omega+\Omega)
G(k',\Omega)G(p+q-k',\omega+\Omega)
T'(p+q,\omega+\Omega).
\end{aligned}
\end{equation}
The conventional RPA (fully self-consistent GW) ignores all the ladder diagrams with short-range interaction 
(i.e., noninteracting pair propagator),
but since we consider tha ladder diagrams with the here, 
the particle number and the energy are nomore conserved and the $f$-sum rule is thus unsatisfied.
Since the ring-diagram approximation is exact in the limit that the kinetic energy is much larger
than the interparticle potential,
the screened long-range interaction (between majority particles) should not be taken into account.
However, while for strong-interacting case,
the self energy of the majority particles in Hartree-Fock approximation (GV approximation) read 
\begin{equation} 
\begin{aligned}
\Sigma_{\uparrow}(q)=\frac{g_{\psi\psi}}{\Omega+i0-\varepsilon_{q\uparrow}-\Sigma_{\uparrow}(0)}
+\int\frac{d^{3}p}{(2\pi)^{3}}\int\frac{d\omega}{2\pi}G^{0}(p)T^{(1)}(p+q,\omega+\Omega),
\end{aligned}
\end{equation}
where $T^{(1)}(p+q,\omega+\Omega)$ is the first order term in $T$-matrix expansion.

When the three-dimensional attractive potential forms a bound state,
the bare coupling can be written as 
\begin{equation} 
\begin{aligned}
\frac{1}{g_{b}}=-\int\frac{d^{3}k}{(2\pi)^{3}}\frac{1}{E_{b}+\varepsilon_{k\uparrow}+\varepsilon_{k\downarrow}+W}.
\end{aligned}
\end{equation}
The binding energy $E_{b}$ usually cannot be tuned in solid state system,
and the center-of-mass momentum of polaron vanishes for large enough $E_{b}$.
$W$ is the bandwidth which decreases with the increasing interaction strength (mainly the local exchange correlation).
An ultra-violet cutoff $\Lambda$ is needed. Although the integral over $k$ shows the bare coupling   
is independent of the momentum,
but indeed we have $g_{b}\propto 1/\Lambda$ and thus $g_{b}\propto 1/{\rm max}(k)^{2}$ (for parabolic systems).
As shown in Fig.11, the interaction strength is logarithmically divergent in the UV cutoff,
and thus the $T$-matrix should also be logarithmically divergent in the UV cutoff.
The bare coupling $g_{b}$ is inversely proportional to the momentum cutoff $\Lambda$.
We can see that the coupling $g_{b}$ depends more on UV cutoff than on binding energy,
which makes the case of $E_{b}\gg|\Sigma|$ becomes possible (e.g., when $\Lambda\rightarrow \infty$),
and that is also the case that the mean-field description as well as the ring-diagram approximation\cite{Sarma S D}
can be well applied,
i.e., when the interaction potential is much weaker than the single particle kinetic energy.

While in the long-wavelength limit ($k\rightarrow 0$), which is the strong-interacting case,
the Thomas-Fermi theory is precise (due to the strong long-wavelength response),
and we have $g_{b}^{-1}\rightarrow 0$, 
$T(p+q,\omega+\Omega)\rightarrow 1/(-\Pi(p+q,\omega+\Omega))$.
$\Pi(p+q,\omega+\Omega)$ is independent of the $g_{b}$ but dependent on the cutoff $\Lambda$.
In this case, the resulting pair propagator is much smaller (Fig.12(a)) and the interaction vertex 
(as well as the self energy) increases exponentially with the impurity momentum $p$ (Fig.12(b)).
And for $p\rightarrow 0$ limit, both the pair propagator and the vertex show
 linear dependence on $p$, which is correct no matter what kinds of momentum-dependence
the $g_{b}$ is.

\section{Appendix.C: Temperature effect and the three-body correlation}

The temperature effect, as shown in Sec.5.,
plays a role in the polaron formation and the many-body instability,
which is widely explored theoretically and experimentally\cite{Onofrio R,Field B,Guenther N E},
However we ignore the effect of electron-phonon coupling in Sec.5 and 
under the restriction of low-temperature limit.
Specially, we note that in low-carrier-concentration limit 
($n=\int^{\mu}_{-\infty}d\omega N_{F}(\omega)D(\omega)$; $D(\omega)$ 
is the density-of-states), as we mentioned in the introduction,
the semiclassical Boltzmann conductivity of dilute metals (like the doped semiconductor)
should be low, 
which reads $\sigma=\frac{e^{2}v_{F}^{2}}{3}\int d\omega D(\omega)\tau(\omega)
[\frac{-\partial N_{F}(\omega)}{\partial \omega}]$
and can has a Drude-like form 
$\sigma=\frac{e^{2}n}{m^{*}}\int d\omega \tau(\omega)
[\frac{-\partial N_{F}(\omega)}{\partial \omega}]$
 if $1/\tau\ll E_{F}$, and in this case
even 
the moderate electron-phonon coupling can leads to the strange metallicity.
Note that the term $[\frac{-\partial N_{F}(\omega)}{\partial \omega}]=\delta(\omega)$ at low-enough temperature
\cite{Werman Y}.
In this section, we will take the electron-phonon coupling (and phonon-mediate attraction)
into account and study a kind of more complex polaron formed by the three-body correlation
(phonon, impurity, and electron-hole pair) as well as its thermodynamics.


We at first discuss the weak coupling adiabatic case, where the 
Migdal-Eliashberg theory is applicable,
 then the (single species) self-energy of the phonon (no matter transverse or longitudinal modes)
can be obtained by the summation of the electronic states (excited by the electron-phonon interaction) with different crystal momenta during the transition
\begin{equation} 
\begin{aligned}
\Sigma_{ph}(q',\Omega_{ph})=\int\frac{d^{2}k}{(2\pi)^{2}}|g_{p,p+q'}|^{2}\frac{N_{F}(\varepsilon_{k+q'})-N_{F}(\varepsilon_{k})}
{\varepsilon_{k+q'}-\varepsilon_{q'}-\Omega_{ph}-i0},
\end{aligned}
\end{equation}
where $g_{p,p+q'}$ is the well known electron-phonon coupling matrix element\cite{Li X,Matthes L,Borysenko K M,Yan J A,Yan J A2}
which is inversely proportional to the atomic mass and the bare phonon frequency $\Omega^{0}_{ph}(q')$, 
just similar to the descriptions in collective Holstein approximation.
We have in perturbational treatment
\begin{equation} 
\begin{aligned}
g_{p,p+q'}=\sqrt{\frac{\hbar}{2M\Omega_{ph}^{0}}}\langle p+q'|\frac{\delta U}{\delta u_{q}}|p\rangle,
\end{aligned}
\end{equation}
where $M$ is the atomic mass of the lattice and $\delta U$ is the variation of the self-consistent Kohn-Sham potential.
$\delta u_{q}$ denotes the tight-binding amplitude.
For acoustic phonon\cite{Bistritzer R},
the dispersion has $\Omega_{ph}^{0}=
v_{s}q'$ where $v_{s}$ is the sound velocity which is related to the 
Bloch-Gruneisen temperature and Debye temperature,
and 
the interaction matrix element can also be related to the chiral relation as
\begin{equation} 
\begin{aligned}
|g_{p,p+q'}|^{2}=
\frac{D^{2}q'^{2}}{2\rho_{M}\Omega_{ph}^{0}}|\langle p+q'|p\rangle|^{2}=
\frac{D^{2}q'}{2\rho_{M}v_{s}}\frac{1+\lambda\lambda'{\rm cos}\theta}{2},
\end{aligned}
\end{equation}
where $D$ is the screened (quasistatic) deformation potential, $\rho_{M}$ is the mass density,
and $\theta$ denotes the angle between $p+q'$ and $p$.
Since here the change of quantum numbers 
is ignored, the phonon self-energy becomes zero in the limit $q'\rightarrow 0$.
We can see that this self-energy is described by the polarization loop (particle-hole bubble)
with the crossing phonon vertex correction.
Note that in Migdal approximation, 
the effect of phonon vertex correction is small and even negligible\cite{Alexandrov A S}
due to the weak electron-phonon coupling strength compared to the electron tunneling
(and thus the leading order scattering theory is valid),
$\varepsilon_{e-ph}(\sim |g_{p,p+q'}|^{2})\ll 3J\Omega_{ph}$ ($3t$ here is the approximated 
bare half bandwidth;
for large polaron, the half bandwidth can also be written as $\frac{3}{2ma^{2}}$\cite{Devreese J T}),
in which case the effective mass of electron ($m^{*}=m(1+\frac{\varepsilon_{e-ph}}{3t\Omega^{0}_{ph}})$) is closes to the rest one.
Note that the above phonon self-energy induced by the propagating electron
is valid only within the Migdal-Eliashberg approximation, 
i.e., the weakly coupled adiabatic limit $\frac{\varepsilon_{e-ph}}{3J\Omega^{0}_{ph}}\ll 1,\ \frac{J}{\Omega^{0}_{ph}(q')}\ll 1$.
(In the insulator phase with large electron-phonon couping, the Migdal-Eliashberg approximation breaks down).
The real part of above phonon self-energy provides the phonon frequency shift while its imaginary part provides
the phonon linewidth (or the population decay\cite{Kohstall C}) since it does not contains the phonon-related vertex correction.
The phonon linewidth is proportional to the electron-phonon coupling matrix element
and inversely proportional to the Fermi velocity of electron,
i.e., the self-energy will induces a higher linewidth in nonadiabatic case,
in other word, the linewidth increases with the increase of doping level.
Note that here the contributions to linewidth from the anharmonic term (inter-phonon interaction)
and the interaction between phonons and the electron-hole pairs are neglected.

As diagrammatically shown in the Fig.1, the boson self-energy loop (phonon)
obtained by many-body diagrammatic method
is comprised of two Fermion propagators (the impurity) and two boson propagators (with external momentum and external frequency) where we consider the current-current correlation here.
The phonon vertex correction reads
\begin{equation} 
\begin{aligned}
\sigma_{j}\Gamma_{ph}=\sigma_{j}-\sum_{q',\Omega_{ph}}
TD_{ph}(q',\Omega_{ph})|g_{k,k+q}|^{2}G_{0}(k+q',\omega+\Omega_{ph})\sigma'_{j}\Gamma'_{ph}G_{0}(k+q+q',\omega+\Omega+\Omega_{ph}),
\end{aligned}
\end{equation}
where $\omega$ is the Fermion frequency and $\Omega/\Omega_{ph}$ is the Bosonic frequency.
In the weak-coupling regiem,
$D_{ph}^{0}(q',\Omega_{ph})=\frac{2\Omega^{0}_{ph}(q')}{\Omega_{ph}^{2}-(\Omega^{0}_{ph}(q'))^{2}}$
is the bare phonon propagator,
which should be replaced by the interacting one when the electron-phonon coupling is strong
(and thus with a larger phonon self-energy):
\begin{equation} 
\begin{aligned}
D_{ph}(q',\Omega_{ph})
=\frac{2\widetilde{\Omega}^{0}_{ph}(q')}{\Omega_{ph}^{2}-(\widetilde{\Omega}^{0}_{ph}(q'))^{2}
-2\widetilde{\Omega}^{0}_{ph}(q')\Sigma_{ph}(q',\Omega_{ph})}
\end{aligned}
\end{equation}
where $\widetilde{\Omega}^{0}_{ph}=\Omega^{0}_{ph}(q')(1-2\frac{\varepsilon_{e-ph}}{J})$ is the renormalized phonon frequency
\cite{Alexandrov A}.

When the polaron-phonon interaction is taken into account,
the self-energy of the impurity induced by electron-phonon interaction reads
\begin{equation} 
\begin{aligned}
\Pi_{e}(p)=&\int\frac{d^{2}q'}{(2\pi)^{2}}|g_{k,k+q'}|^{2}
D_{ph}(q',\Omega_{ph})G_{0}(p+q',\omega+\Omega_{ph})\\
=&\int\frac{d^{2}q'}{(2\pi)^{2}}|g_{p,p+q'}|^{2}
(\frac{1+N_{B}(\Omega^{0}_{ph}(q'))}{\varepsilon_{p+q'}-\varepsilon_{p}+\Omega^{0}_{ph}(q')+i0}
 +\frac{N_{B}(\Omega^{0}_{ph}(q'))}{\varepsilon_{p+q'}-\varepsilon_{p}-\Omega^{0}_{ph}(q')-i0}),
\end{aligned}
\end{equation}
where the analytical continuation $i\Omega\rightarrow \Omega+i\eta^{+}$ is used
and the first and second term correspondto the emission and absorption of a phonon with mometum $q'$,
as persistently happen during the coherent propagation of the phonon.
The hopping containing the effect of phonons (Fr{\"o}hlich interaction) reads
$J'=J{\rm exp}[\sum_{q'}\frac{|g_{p,p+q'}|}{\sqrt{N}}(c_{i}^{\dag}c_{i}d_{q'}-c_{j}^{\dag}c_{j}d_{q'})-h.c.]$,
and the electron inelastic scattering rate reads
\begin{equation} 
\begin{aligned}
\frac{1}{\tau_{e}}=\frac{2\pi}{\hbar}\sum_{q'}|g_{p,p+q'}|^{2}[(N_{B}(\Omega^{0}_{ph}(q'))+1)\delta(\varepsilon_{k-q'}-\varepsilon_{k}+\Omega^{0}_{ph}(q'))
+N_{B}(\Omega^{0}_{ph}(q'))\delta(\varepsilon_{k+q'}-\varepsilon_{k}-\Omega^{0}_{ph}(q'))].
\end{aligned}
\end{equation}
Note that the first term within the bracket contains the effect of spontaneous emission
(of a phonon) even ar zero temperature\cite{Gunst T}.
Apparently, at zero-temperature, the scattering rate (relaxation)
is related to the selection rule 
$\varepsilon_{p-q'}-\varepsilon_{p}=-\Omega^{0}_{ph}(q')$,
and it is coupled to the self-induced lattice polarization here.

Consider the simplified three-body ansatz,
the many-body Hamiltonian of complex polaron state which
contains the interactions between impurity and particle-hole excitation and between impurity and phonon 
(Peierls electron-phonon interaction\cite{Marchand D J J}) reads
\begin{equation} 
\begin{aligned}
H=&\sum_{k}\varepsilon_{k\uparrow}c_{k\uparrow}^{\dag}c_{k\uparrow}+\sum_{p}\varepsilon_{p\downarrow}c_{p\downarrow}^{\dag}c_{p\downarrow}
+\frac{1}{N}\sum_{k,p,q}g_{q}c_{p-q\downarrow}^{\dag}c_{k+q\uparrow}^{\dag}c_{k\uparrow}c_{p\downarrow}\\
&+\sum_{q'}\Omega_{ph}^{0}(q')b_{q'}^{\dag}b_{q}
+\frac{1}{\sqrt{N}}\sum_{k,q',\sigma}|g_{k+q',k}|
c^{\dag}_{p+q',\sigma}c_{p,\sigma}(b^{\dag}_{-q'}+b_{q'}),
\end{aligned}
\end{equation}
where $N=S/s_{0}$ is the total number of unit cell where $S$ is the total area and $s_{0}$ is the area of unit cell.
$g_{q}^{-1}=-\sum_{k}[E_{b}+\varepsilon_{k\uparrow}+\varepsilon_{k\downarrow}+W]^{-1}$.
In last term we use the Frohlich model (or the Fourier transformed SSH model),
where the interaction vertex is symmetry with respect to the sign of interaction,

Then the ground state wave function reads (similar to Eq.(13))
\begin{equation} 
\begin{aligned}
|\psi\rangle_{CP}=&(\phi'_{0}c^{\dag}_{p\downarrow}+\sum_{k>k_{F},q<k_{F}}\phi'_{kq}c_{p+q-k\downarrow}^{\dag}c_{k\uparrow}^{\dag}c_{q\uparrow}
+
\sum_{|q'|\le q_{D}}\phi_{q'}
b_{q'}^{\dag}c_{p-q'\downarrow}^{\dag}\\
&+
\frac{1}{\sqrt{N}}
\sum_{k>k_{F},q<k_{F},|q'|\le q_{D}}\phi_{kqq'}
c_{p+q-k-q'\downarrow}^{\dag}c_{k\uparrow}^{\dag}c_{q\uparrow}b_{q'}^{\dag})|0\rangle
,
\end{aligned}
\end{equation}
where we donot consider the hole excitations excited from the thermal cloud here.
Here $q_{D}$ denotes the Debye radiu and $1/N$ can be treated as the average phonon number.
For the complex polaron,
by minimizing $\langle \psi|E-H|\psi\rangle$,
we have (consider the case that emits one phonon)
\begin{equation} 
\begin{aligned}
\langle \psi_{CP}|E_{\downarrow}-H|\psi\rangle_{CP}=&
E_{\downarrow}(|\phi'_{0}|^{2}+\sum_{kq}|\phi_{kq}|^{2}+\sum_{q'}|\phi_{q'}|^{2}+\sum_{kqq'}|\phi_{kqq'}|^{2})\\
&  -[|\phi'_{0}|^{2}\varepsilon_{p\downarrow}+\sum_{kq}(\varepsilon_{p+q-k\downarrow}+\varepsilon_{k\uparrow}-\varepsilon_{q\downarrow})|\phi_{kq}|^{2}\\
&    +\sum_{q'}(\varepsilon_{p-q'\downarrow}+\varepsilon_{-q'})|\phi_{q'}|^{2}
    +\sum_{kqq'}(\varepsilon_{p+q-k-q'}+\varepsilon_{k\uparrow}-\varepsilon_{q\uparrow}+\Omega_{ph}^{0}(q'))|\phi_{kqq'}|^{2}\\
&    +|\phi'_{0}|^{2}\sum_{q}g_{q}
    +\sum_{kq}(\phi^{'*}_{0}\phi_{kq}g_{k-q}+c.c.)
    +\sum_{q'}(\phi^{'*}_{0}\phi_{q'}g_{q'}+c.c.)\\
&    +\sum_{kqq'}(\phi^{'*}_{0}\phi_{kqq'}g_{k-q+q'}+c.c.)\\
&    +\sum_{kqq'}(\phi^{'*}_{kq}\phi_{q'}g_{k-q+q'}+c.c.)
    +\sum_{kqq'}(\phi^{'*}_{kq}\phi_{kqq'}g_{q'}+c.c.)
    +\sum_{kqq'}(\phi^{'*}_{q'}\phi_{kqq'}g_{k-q}+c.c.)
].
\end{aligned}
\end{equation}
Here we ignore the change of the momentum of electron-hole pair.
Then for $\frac{\partial \langle\psi_{CP}|H|\psi_{CP}\rangle}{\partial \phi}=0$,
we write
\begin{equation} 
\begin{aligned}
\varepsilon_{p\downarrow}\phi_{0}+\phi_{0}\sum_{q}g_{q}+\sum_{kq}\phi_{kq}g_{k-q}+\sum_{q'}\phi_{q'}g_{q'}+\sum_{kqq'}\phi_{kqq'}g_{k-q+q'}=E\phi_{0},\\
(\varepsilon_{p+q-k\downarrow}+\varepsilon_{k\uparrow}-\varepsilon_{q\downarrow})\phi_{kq}+\phi_{0}g_{k-q}+\sum_{kqq'}\phi_{q'}g_{k-q+q'}+\sum_{kqq'}\phi_{kqq'}g_{q'}=E\phi_{kq},\\
(\varepsilon_{p-q'\downarrow}+\varepsilon_{q'})\phi_{q'}+\phi_{0}g_{q'}+\sum_{kqq'}\phi_{kq}g_{k-q+q'}+\sum_{kqq'}\phi_{kqq'}g_{k-q}=E\phi_{q'},\\
(\varepsilon_{p+q-k-q'}+\varepsilon_{k\uparrow}-\varepsilon_{q\uparrow}+\Omega_{ph}^{0}(q'))\phi_{kqq'}
+\phi_{0}g_{k-q+q'}+\sum_{kqq'}\phi_{kq}g_{q'}+\sum_{kqq'}\phi_{q'}g_{k-q}
=E\phi_{kqq'},
\end{aligned}
\end{equation}
The variational paramters are obtained by the minimization as
\begin{equation} 
\begin{aligned}
\phi_{0}=&\frac{\frac{g}{N}\sum_{q}\chi_{q}}{E_{\downarrow}-\varepsilon_{p\downarrow}},\\
\phi_{kq}=&\frac{\frac{g}{N}\chi_{q}}
           {E_{\downarrow}-(\varepsilon_{p+k-q\downarrow}+\varepsilon_{k\uparrow}-\varepsilon_{q\uparrow})},\\
\phi_{q'}=&\frac{\frac{g}{N}\sum_{q}\chi_{q}}
           {E_{\downarrow}-(\varepsilon_{p-q'\downarrow}+\Omega_{ph}^{0}(q'))},\\
\phi_{kqq'}=&\frac{\frac{g}{N}\chi_{q}}
           {E_{\downarrow}-(\varepsilon_{p+k-q-q'\downarrow}+\varepsilon_{k\uparrow}-\varepsilon_{q\uparrow}+\Omega_{ph}^{0}(q'))},\\
\chi_{q}=&\psi_{0}+\sum_{k}\psi_{kq}+\sum_{q'}\psi_{q'}+\sum_{kq'}\psi_{kqq'},\\
\end{aligned}
\end{equation}
where the self-consistent energy term $(E-\varepsilon_{p\downarrow})$ reads
\begin{equation} 
\begin{aligned}
E-\varepsilon_{p\downarrow}=&\frac{1}{N}\sum_{q<k_{F}}\left[g^{-1}-\frac{1}{N}\sum_{k=k_{F}}^{\Lambda}\frac{1}{E+i0
-(\varepsilon_{p+k-q-q'\downarrow}+\varepsilon_{k\uparrow}-\varepsilon_{q\uparrow}+\Omega_{ph}^{0}(q'))}
\right]^{-1},
\end{aligned}
\end{equation}
which equals the binding energy when $p=0$.
For small electron chemical potential, we define
\begin{equation} 
\begin{aligned}
g^{-1}(\Lambda)=&-\frac{1}{N}\sum_{k,q'}^{\Lambda}\frac{1}{E_{b}+\varepsilon_{k\uparrow}+\varepsilon_{k\downarrow}+\Omega_{ph}^{0}(q')+W}.
\end{aligned}
\end{equation}
Unlike the renormalized coupling parameter in Eq.(6),
we can see that this coupling parameter depends on both the binding-energy $E_{b}$ and momentum (ultraviolet) cutoff $\Lambda$.
and the logarithmically diverges of $g^{-1}(\Lambda)=-\infty$ can be seen
($g(\Lambda)\rightarrow 0^{-}$) when $\Lambda\rightarrow\infty$ (where the quantum correction vanishes).
The normalization condition is ${}_{CP}\langle \psi|\psi\rangle_{CP}=
|\phi_{0}|^{2}+\sum_{k>k_{F},q<k_{F}}|\phi_{kq}|^{2}+N_{ph}\sum_{q'\le q_{D}}|\phi_{q'}|^{2}+N_{ph}
\sum_{k>k_{F},q<k_{F},|q'|\le q_{D}}|\phi_{kqq'}|^{2}=1$.
Note that such a complex polaron cannot be
found in a superfluid bath where the propagating impurity will not creates the electron-hole excitations.
The binding energy $E_{b}$ here is positive for attractive potential, i.e., the Fermi polaron here
with negative $g_{q}$,
thus the energy of such a bound state must be negative $-E_{b}<0$,
which corresponds to the negative self-energy of the attractive Fermi polaron.

If there only exist two possibilities: isolate impurity, or 
interacting with both the electron-hole pair and the phonons simultaneously.
Then the variational wave function reads (here consider the one particle-hole pair and multi-phonon case)
\begin{equation} 
\begin{aligned}
|\psi\rangle_{CP}=&(\phi'_{0}c^{\dag}_{p\downarrow}
+\sqrt{N_{ph}}
\sum_{k>k_{F},q<k_{F},|q'|\le q_{D}}\phi_{kqq'}
c_{p+q-k-q'\downarrow}^{\dag}c_{k\uparrow}^{\dag}c_{q\uparrow}b_{q'}^{\dag})|0\rangle
,
\end{aligned}
\end{equation}
where the factor $\sqrt{N_{ph}}$ is 
created by the $b_{-q'}$ acting on the unperturbed state $|0\rangle$.
Unlike the anharmonic phonon modes induced by the high intensity laser\cite{Zijlstra E S}
where the interphonons coupling can not be ignored,
we ignore the interphonons couping here as the modes are created by the self-induced lattice polarization.
The variational parameters are obtained by the minimization as
\begin{equation} 
\begin{aligned}
\phi_{0}=&\frac{\frac{gN_{ph}}{N}\sum_{q}\chi_{q}}{E_{\downarrow}-\varepsilon_{p\downarrow}}
=\frac{gn_{ph}\sum_{q}\chi_{q}}{E_{\downarrow}-\varepsilon_{p\downarrow}},\\
\phi_{kqq'}=&\frac{\frac{g}{N}\chi_{q}}
           {E_{\downarrow}-(\varepsilon_{p+k-q-q'\downarrow}+\varepsilon_{k\uparrow}-\varepsilon_{q\uparrow}+\Omega_{ph}^{0}(q'))},\\
\chi_{q}=&\psi_{0}+\sum_{kq'}\psi_{kqq'},\\
\end{aligned}
\end{equation}
where $n_{ph}$ denotes the phonon density.
The normalization condition is ${}_{CP}\langle \psi|\psi\rangle_{CP}=
|\phi_{0}|^{2}+N_{ph}
\sum_{k>k_{F},q<k_{F},|q'|\le q_{D}}|\phi_{kqq'}|^{2}=1$.
Here the momentum distribution of polaron is shifted from the $\delta_{p}$-function
by the electron-phonon coupling term $|\phi_{kqq'}|^{2}$.
Through the above procedure,
we can obtain
\begin{equation} 
\begin{aligned}
\frac{\phi_{kqq'}}{\phi_{0}}=&\frac{g\frac{1}{N}\chi_{q}(E-\varepsilon_{p\downarrow})}{g\frac{N_{ph}}{N}\sum_{q}\chi_{q}
(E-(\varepsilon_{p+k-q-q'\downarrow}+\varepsilon_{k\uparrow}-\varepsilon_{q\uparrow}+\Omega_{ph}^{0}(q')))}\\
=&\frac{1}{N_{ph}N}\frac{E-\varepsilon_{p\downarrow}}
{E-(\varepsilon_{p+k-q-q'\downarrow}+\varepsilon_{k\uparrow}-\varepsilon_{q\uparrow}+\Omega_{ph}^{0}(q')))},
\end{aligned}
\end{equation}
where the energy is
\begin{equation} 
\begin{aligned}
E-\varepsilon_{p\downarrow}=&\frac{1}{N}\sum_{q<k_{F}}\left[g^{-1}-\frac{1}{N}\sum_{k=k_{F}}^{\Lambda}\frac{1}{E+i0
-(\varepsilon_{p+k-q-q'\downarrow}+\varepsilon_{k\uparrow}-\varepsilon_{q\uparrow}+N_{ph}\Omega_{ph}^{0}(q'))}
\right]^{-1},\\
=&\frac{1}{N}\sum_{q<k_{F}}\left[
-\frac{1}{N}\sum_{k,q'}^{\Lambda}\frac{1}{E_{b}+\varepsilon_{k\uparrow}+\varepsilon_{k\downarrow}+N_{ph}\Omega_{ph}^{0}(q')+W}\right.\\
&\left.-\frac{1}{N}\sum_{k=k_{F},q'}^{\Lambda}\frac{1}{E+i0
-(\varepsilon_{p+k-q-q'\downarrow}+\varepsilon_{k\uparrow}-\varepsilon_{q\uparrow}+N_{ph}\Omega_{ph}^{0}(q'))}
\right]^{-1}.
\end{aligned}
\end{equation}
Note that for the majority component,
the eigenenergy should contains the chemical potential,
e.g., in above equation we have $\varepsilon_{k\uparrow}=\frac{k^{2}}{2m_{\uparrow}}-\mu_{\uparrow}$
and $\varepsilon_{q\uparrow}=\frac{q^{2}}{2m_{\uparrow}}+\mu_{\uparrow}$,
i.e., the pair energy is measured from $2\mu_{\uparrow}$.

Consider the electron-phonon coupling, where the phonon can be treated as a thermal excited Boson
in the diagrammatic expansion of the self-energy, as shown in Fig.13,
where $\Sigma_{2}(p,\omega)$ reads
\begin{equation} 
\begin{aligned}
\Sigma_{2}(p,\omega)=\sum_{q'}T_{2}G(q')
           =\sum_{q'}\frac{N_{B}G(q')}{g_{k+q',k}^{-1}-\frac{1}{\omega+i0-(\varepsilon_{p-q'}+\varepsilon_{q'})}},
\end{aligned}
\end{equation}
and
$\Sigma_{3}(p,\omega)$ reads
\begin{equation} 
\begin{aligned}
\Sigma_{3}(p,\omega)=\sum_{q'}T_{3}G(q')
=\sum_{q'}\frac{N_{B}G(q')}{g_{k+q',k}^{-1}-\frac{1}{\omega+i0-(\varepsilon_{p+q-k-q'\downarrow}+\varepsilon_{q'}+\varepsilon_{k\uparrow}-\varepsilon_{q\uparrow})}}.
\end{aligned}
\end{equation}
Obviously, the self-energy $\Sigma_{2}(p,\omega)$ is induced by the interaction between phonon and the bare impurity
while the self-energy $\Sigma_{3}(p,\omega)$ is 
induced by the interaction between phonon and the dressed impurity (Fermi polaron).
Note that due to the existence of Bose distribution function
in numerator, these two self-enegy terms vanish at zero-temperature.
The final self-energy can be written as
\begin{equation} 
\begin{aligned}
\Sigma(p,\omega)=&
\sum_{q,\Omega}T(p+q,\omega+\Omega)G_{0}(q,\Omega)
+\sum_{q',k,q,\Omega_{ph}}T_{2,3}(p+q-k-q',\omega+\Omega+\Omega_{ph})G_{0}(q',\Omega_{ph})\\
&+\sum_{q',k,q,\Omega_{ph}}
 \frac{N_{F}}{T^{-1}(p+q,\omega+\Omega)-\frac{N_{B}}{T_{2,3}^{-1}(p+q-k-q',\omega+\Omega+\Omega_{ph})}}G_{0}(q-q',\Omega+\Omega_{ph}),
\end{aligned}
\end{equation}

\end{large}
\renewcommand\refname{References}

\clearpage

\end{document}